\PassOptionsToPackage{unicode}{hyperref}
\PassOptionsToPackage{hyphens}{url}
\documentclass[
]{article}
\usepackage{amsmath,amssymb}
\usepackage{lmodern}
\usepackage{iftex}
\ifPDFTeX
  \usepackage[T1]{fontenc}
  \usepackage[utf8]{inputenc}
  \usepackage{textcomp} 
\else 
  \usepackage{unicode-math}
  \defaultfontfeatures{Scale=MatchLowercase}
  \defaultfontfeatures[\rmfamily]{Ligatures=TeX,Scale=1}
\fi
\IfFileExists{upquote.sty}{\usepackage{upquote}}{}
\IfFileExists{microtype.sty}{
  \usepackage[]{microtype}
  \UseMicrotypeSet[protrusion]{basicmath} 
}{}
\makeatletter
\@ifundefined{KOMAClassName}{
  \IfFileExists{parskip.sty}{%
    \usepackage{parskip}
  }{
    \setlength{\parindent}{0pt}
    \setlength{\parskip}{6pt plus 2pt minus 1pt}}
}{
  \KOMAoptions{parskip=half}}
\makeatother
\usepackage{xcolor}
\usepackage[margin=1in]{geometry}
\usepackage{color}
\usepackage{fancyvrb}

\DefineVerbatimEnvironment{Highlighting}{Verbatim}{commandchars=\\\{\}}
\usepackage{framed}
\definecolor{shadecolor}{RGB}{248,248,248}
\newenvironment{Shaded}{\begin{snugshade}}{\end{snugshade}}

\newcommand{\AttributeTok}[1]{\textcolor[rgb]{0.77,0.63,0.00}{#1}}

\newcommand{\CommentTok}[1]{\textcolor[rgb]{0.56,0.35,0.01}{\textit{#1}}}

\newcommand{\ConstantTok}[1]{\textcolor[rgb]{0.00,0.00,0.00}{#1}}
\newcommand{\ControlFlowTok}[1]{\textcolor[rgb]{0.13,0.29,0.53}{\textbf{#1}}}

\newcommand{\DecValTok}[1]{\textcolor[rgb]{0.00,0.00,0.81}{#1}}
\newcommand{\DocumentationTok}[1]{\textcolor[rgb]{0.56,0.35,0.01}{\textbf{\textit{#1}}}}

\newcommand{\FloatTok}[1]{\textcolor[rgb]{0.00,0.00,0.81}{#1}}
\newcommand{\FunctionTok}[1]{\textcolor[rgb]{0.00,0.00,0.00}{#1}}

\newcommand{\NormalTok}[1]{#1}

\newcommand{\OtherTok}[1]{\textcolor[rgb]{0.56,0.35,0.01}{#1}}

\newcommand{\SpecialCharTok}[1]{\textcolor[rgb]{0.00,0.00,0.00}{#1}}

\newcommand{\StringTok}[1]{\textcolor[rgb]{0.31,0.60,0.02}{#1}}

\usepackage{graphicx}
\makeatletter
\def\maxwidth{\ifdim\Gin@nat@width>\linewidth\linewidth\else\Gin@nat@width\fi}
\def\maxheight{\ifdim\Gin@nat@height>\textheight\textheight\else\Gin@nat@height\fi}
\makeatother
\setkeys{Gin}{width=\maxwidth,height=\maxheight,keepaspectratio}
\makeatletter
\def\fps@figure{htbp}
\makeatother
\providecommand{\tightlist}{%
  \setlength{\itemsep}{0pt}\setlength{\parskip}{0pt}}
\newlength{\cslhangindent}
\newlength{\csllabelwidth}
\newlength{\cslentryspacingunit} 
\newenvironment{CSLReferences}[2] 
 {
  \setlength{\parindent}{0pt}
  \ifodd #1
  \let\oldpar\par
  \def\par{\hangindent=\cslhangindent\oldpar}
  \fi
  \setlength{\parskip}{#2\cslentryspacingunit}
 }%
 {}
\usepackage{calc}

\ifLuaTeX
  \usepackage{selnolig}  
\fi
\IfFileExists{bookmark.sty}{\usepackage{bookmark}}{\usepackage{hyperref}}
\IfFileExists{xurl.sty}{\usepackage{xurl}}{} 
\hypersetup{
  pdftitle={Chapter 10 - Quantitative models of Discounting},
  pdfauthor={Christopher T. Franck},
  hidelinks,
  pdfcreator={LaTeX via pandoc}}

\title{Chapter 10 - Quantitative Models of Discounting}
\author{Christopher T. Franck, Department of Statistics, Virginia Tech}
\date{2024-08-07}

\begin{document}
\maketitle

\begin{center}
From the forthcoming publication Operant Behavioral Economics published by Elsevier
\end{center}

\hypertarget{overview-of-the-chapter}{%
\subsection{Overview of the chapter}\label{overview-of-the-chapter}}

The ability to confidently fit models to indifference point data
requires expertise in behavioral analytic concepts and statistical
modeling. Since many readers of this book may have more primary training
on the behavioral side of things, the purpose of this chapter is to
de-mystify the process of fitting discounting models to observed data.
There will be a special emphasis on fully reproducible code that allows
the reader to execute the analyses described herein. The chapter
features instructive prose, a real world data set, and snippets of R
code which will enable the interested reader to practice analyzing data.

This chapter provides a tutorial that the reader can follow towards
analyzing discounting data. Previous chapters have already described the
breadth of outcomes associated with discounting (Odum et al. 2020) and
other background information (Odum 2011). We focus on delay discounting,
where indifference points describe the value of a delayed reward that a
participant would be willing to accept in order to have the reward
immediately for a variety of delays. This chapter describes the
two-stage approach to analyzing discounting data, as this is the
simplest approach that is also statistically defensible. The two-stage
approach first quantifies the discounting rate of each participant
individually, and second stage analyzes these rates as a function of
relevant variables (e.g., between smokers and non-smokers).

We use the R software (R Core Team 2022) to illustrate the analyses
presented in this chapter. While several software choices are available,
we choose R because it is a free and robust statistical modeling
package. Especially compared to point-and-click menu driven software,
R's programmatic approach to data analysis enables more control,
creativity, and naturally facilitates reproducibility and transparency
since the code that conducts an analysis can be archived and re-run
later and/or by different users. Despite this chapter's commitment to R,
don't conflate the data-analytic approach with the tools used to execute
the analysis. Other software packages exist that are just as capable of
organizing and visualizing data, fitting models, and conducting other
statistical analyses.

Several approaches have been used to analyze multi-subject behavioral
economic data. An overview of these approaches in the context of demand
can be found here (Kaplan et al. 2021). Much of the terminology in this
paper (especially the terms defined in Table 1 of that paper) is
extremely relevant for the analysis of discounting data. In addition, if
the reader is interested in learning about quantifying behavioral
economic demand data, please see Chapter 5 - Quantitative Models of
Demand in this book.

The study of discounting is vast and no single chapter can cover
everything. Once the skills in this chapter are familiar, the reader may
wish to next consider: Additional models for delay discounting
(McKerchar et al. 2009), Model selection including the tension between
model fit and theoretical appeal of models (Franck et al. 2015),
probability discounting (Rachlin et al. 1986; Rachlin, Raineri, and
Cross 1991; Killeen 2023), hierarchical modeling (M. E. Young 2017;
Chávez et al. 2017), useful cross-model metrics of delay discounting
(e.g., Effective delay 50 (Yoon and Higgins 2008; Franck et al. 2015),
area under the fitted curve (Gilroy and Hantula 2018)) , and Bayesian
statistical approaches for delay discounting (Franck et al. 2019).

\hypertarget{how-to-use-this-chapter}{%
\subsection{How to use this chapter}\label{how-to-use-this-chapter}}

The remainder of the chapter is a tutorial. The next step is to download
R (\url{https://www.r-project.org/}) and the free interface program
RStudio (\url{https://posit.co/downloads/}). You can easily find a three
minute long video on the internet that explains how to download and
install R and RStudio for either PC or Mac. Once that is complete, come
along on the journey by running the embedded code as you encounter it in
the chapter.

We alternate between descriptive prose, lines of R code that run
statistical analyses, the results of those analyses, and a description
of those results. We reserve \textbf{bold} font for terms that, if
unfamiliar, can be Googled by the reader to better understand the
context of what is being presented. This chapter is designed to be a
pathway, but on your journey, stopping to read more about related
concepts is the equivalent of sight-seeing. We conclude the chapter with
a series of exercises to give the reader the opportunity to flex their
new knowledge. Treating these exercises like homework might even make
you feel like a kid again.

Please note that this chapter is a broad overview of many topics and is
not organized to reflect a typical research analysis pipeline. Many
planning steps and other considerations must take place before
responsible data analysis, and these issues are described further later
in the chapter in the \emph{Contemporary issues in statistical practice}
and \emph{The role of planning in scientific investigation} Sections.

\hypertarget{r-and-rstudio}{%
\subsection{R and RStudio}\label{r-and-rstudio}}

R is a free programming language, and RStudio (\url{https://posit.co/})
is a convenient graphical user interface for R with a free individual
use license. Once R and RStudio are installed on your machine, you open
RStudio and it automatically connects with R. All R programming, running
of code, and other analysis happens within RStudio.

\begin{figure}
\includegraphics[width=0.9\linewidth]{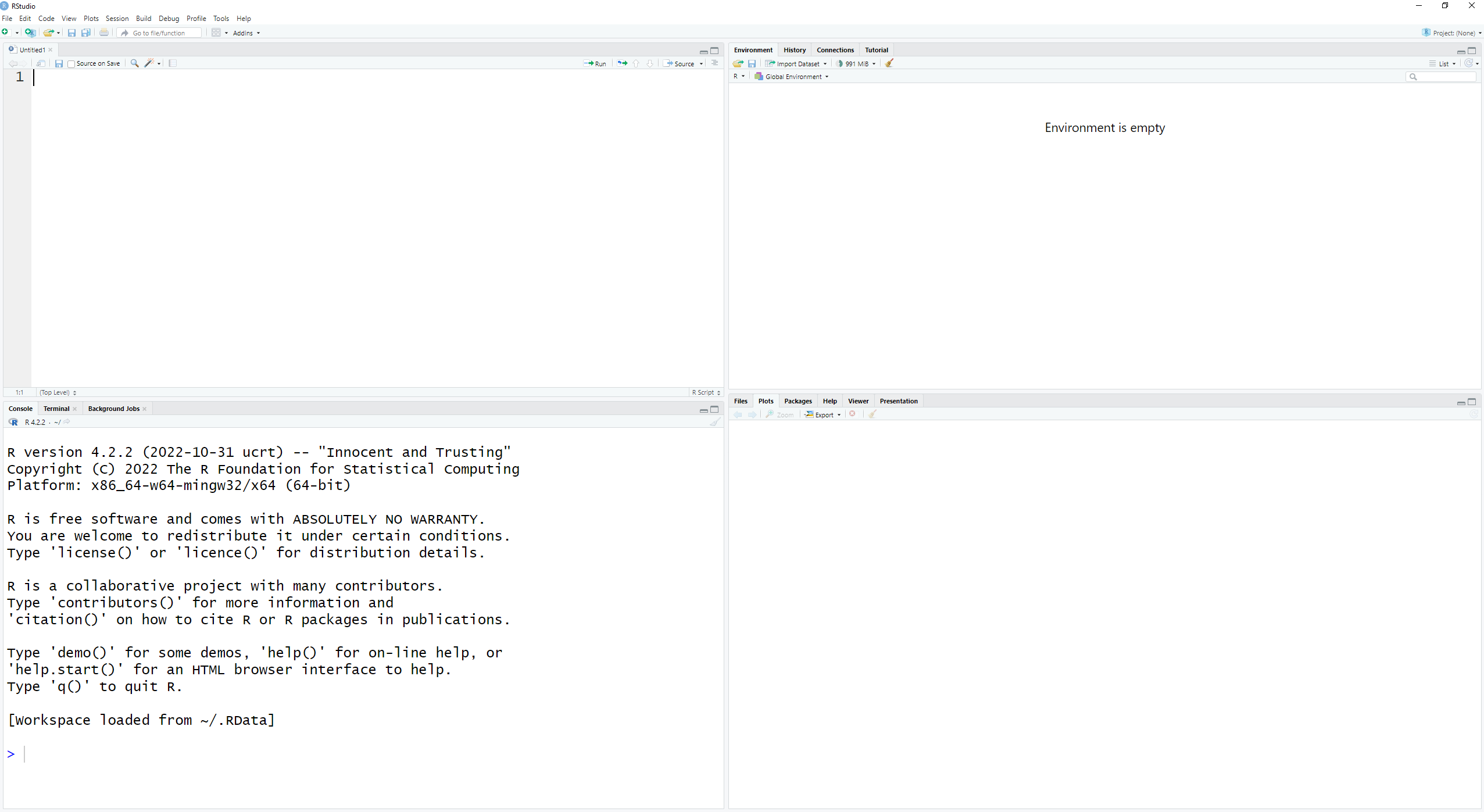} \caption{Rstudio interface. Top left pane is the script, bottom left pane is the console, top right pane shows the (currently empty) environment, and bottom right pane can show help files and data visualizations.}\label{fig:figurename}
\end{figure}

The above Figure depicts the Rstudio interface. The top left pane, which
is blank, is the \textbf{script}. This is where the user (you) writes R
code. A good way to follow along with the demos below is to copy and
paste the code from this chapter into the R script, modify as necessary
(e.g.~change file paths), then run the code. On PC, the line of code
with the cursor can be run by pressing ``Ctrl+Enter''. On Mac, press
``command+return'' to run the current line of code. On both platforms,
you can run multiple lines of code by highlighting those lines you wish
to run before using the run commands just described. R is case
sensitive.

The bottom left panel is the \textbf{console}. When lines of code are
run from the script, those lines appear in the console followed by R's
reaction to those lines. To familiarize yourself with this, first treat
the console as a simple calculator. Type ``2+2'' directly into the
console, press ``Enter'' (``return'' on Mac), and note that R comes back
with ``4''. Now write ``2+2'' in the script, press ``Ctrl+Enter''
(``command+return'' on Mac), and you will see the same results appear in
the console. You will frequently find yourself in a workflow loop in
which you enter commands into the script, submit them to the console,
then assess the outcome of those commands by checking the console for
results and any warning or error messages (which you should Google to
help you understand and resolve).

The top right panel of RStudio shows the \textbf{environment} by
default. The environment lists each \textbf{object} the user has defined
along with some basic information about those objects. Objects include
data the user may have imported, any functions they have written, any
variables that are stored, etc. The bottom right panel can be used to
show graphical output and help files. Keep your eyes on the top right
and bottom right panels as you run the code below to get a feel for the
role these panels play.

We could fill the rest of the chapter describing R and Rstudio and not
run out of things to talk about. But the purpose of this chapter is to
describe the analysis of discounting data. The reader is encouraged to
liberally use other available R resources alongside this chapter as
necessary, but for now, we turn our attention towards the analysis of
discounting data.

\hypertarget{use-r-to-import-and-examine-data}{%
\subsection{Use R to import and examine
data}\label{use-r-to-import-and-examine-data}}

Let us now import the study data. The data analyzed in this chapter were
gathered as part of a larger study (Traxler et al. 2022). These data are
currently available from the author's website. Before replicating the
analysis below, be sure you have downloaded the data and stored it in
the folder of your choice on your computer. The first line of code below
indicates to R which folder stores the data. You will need to modify
this line to reflect the data file's location on your own machine. The
second line defines an object called \emph{dat}, which contains the full
set of data used in this chapter. Copy-paste the following code into an
R script, change the ``setwd'' directory to reflect the location of the
data on your own machine, and run the code.

\begin{Shaded}
\begin{Highlighting}[]
\FunctionTok{setwd}\NormalTok{(}\StringTok{"C:/Dropbox/Discounting book chapter/Rmarkdown"}\NormalTok{) }\CommentTok{\#set working directory}
\NormalTok{dat}\OtherTok{\textless{}{-}}\FunctionTok{read.csv}\NormalTok{(}\StringTok{\textquotesingle{}Traxler 2022.csv\textquotesingle{}}\NormalTok{) }\CommentTok{\#define an object "dat" that is the data}
\end{Highlighting}
\end{Shaded}

The pound sign ``\#'' is used in R to denote \textbf{comments}. Comments
are not evaluated by R, which makes them useful for humans to write
messages to each other within the code (as seen above), and also to
indicate to R if you wish to omit certain lines of code from being
evaluated. This second use is important for \textbf{debugging}.
Essentially all computer programming languages include the ability to
write comments, and thus commenting is a general programming concept.

It is important to examine the data post-import to make sure that the
file was read correctly. You can examine the full set of data by
clicking the ``dat'' entry in the environment pane (top right), or
running the following command in the console:\texttt{View(dat)}. This
will create a tab in the top left pane that shows the study data in a
spreadsheet.

Click the ``dat'' entry in the environment pane now, and confirm you are
able to observe the following. This data set includes as columns id
number, Age (years), gender (``Male'' or ``Female''), and smoking status
as ``smoke\_cigs'' (``Yes'' or ``No''). The data set also includes
indifference points gathered at six delays using hypothetical delay
discounting titrating questionnaire. The delays are one day, one week,
one month, three months, one year, and five years. These indifference
points are labeled y1, y7, y30, y90, y365, and y1825, respectively. Note
that these indifference points are expressed on a scale between zero and
one, as the raw indifference points have already been dived by the
larger later amount in the production of this data set. An attention
check question ``ddattend'' was asked to determine which participants
would choose ``\$0.00 now'' versus ``\$100.00 in 1 day''. Choosing no
money over one hundred dollars tomorrow reflects a participant who
didn't understand the task, was not being attentive to their answers, or
for other reasons is exhibiting irrational patterns of valuation. The
data also include indifference points that violate the Johnson \& Bickel
criteria (Johnson and Bickel 2008), and the presence of these violations
is included in the `JBviol' column. Additional description of attention
checks and non-systematic discounting patterns appears in the next
subsection on first stage analyses. Each row in this spreadsheet
contains the data from a single research participant.

There are two lines of code below that offer additional information
about the data. The first line uses the ``class'' function to indicate
that the data object is stored as a \textbf{data frame}. Data frames are
the typical data format for rectangular data objects in R. Data frames
store variables in the columns and data records (i.e., participant data)
in the rows. Data frames can can include both numeric or character
variables, e.g.~our data has character variable ``gender'' which takes
the values ``Male'' or ``Female'', and numeric data for age and many
others. The \emph{class} function is used on R objects in general to
reveal the \textbf{type} of object. R is able to recognize different
object types, including \textbf{character variables}, \textbf{numeric
variables}, \textbf{vectors of numbers}, \textbf{matrices}, \textbf{data
frames}, lists of different types of objects, models, and many more. A
good first debugging step when faced with warnings and error messages is
to check the type of objects being used.

The second line uses the ``dim'' function to reveal that there are 106
rows (participants) and 13 columns (variables) in the data set.

\begin{Shaded}
\begin{Highlighting}[]
\FunctionTok{class}\NormalTok{(dat)}
\end{Highlighting}
\end{Shaded}

\begin{verbatim}
## [1] "data.frame"
\end{verbatim}

\begin{Shaded}
\begin{Highlighting}[]
\FunctionTok{dim}\NormalTok{(dat)}
\end{Highlighting}
\end{Shaded}

\begin{verbatim}
## [1] 106  13
\end{verbatim}

These data are considered to be in \textbf{wide} format, since the
identifier for participant id does not repeat in different rows. Each
row corresponds to the full data available for a single participant. The
first row, for example, tells us the age, gender, smoking status,
indifference points, and attention and data quality checks for the first
participant. Data in which the same participant ID occurs on multiple
lines (e.g., if each participant completes assessments multiple times)
are said to be in \textbf{long} format. Choice of wide versus tall
format is frequently made based on the software tools the analyst plans
to use, and R functions exist to readily switch between formats.
Recognizing the data format is an important early step since it enables
the researcher to know how to write the code to conduct the analysis.

\textbf{Stage 1 analyses}

The first stage of a two-stage analysis involves quantifying the rate of
discounting for each participant on the basis of their indifference
points. To accomplish this, we first analyze a single participant's
indifference point data. This includes graphically plotting their
indifference points by delay, then fitting a discounting model, adding
the model fit to the plot, and finally extracting the discounting rate
for subsequent analysis in stage two. Don't worry about the fact that we
have to do this 106 times. Once we have the first subject's Stage 1
analysis complete, we will show how to tweak and embed that code in a
\textbf{loop} that atomically and near-instantaneously conducts the same
analysis for the remaining 105 participants and stores the results in a
format convenient for Stage 2 analysis.

Now we plot the first participant's data. Note the ``\textless-'' is an
assignment statement used to define new objects (see how it looks like a
little arrow). So the first line creates a new object called ``i'' which
takes the value ``1''.

\begin{Shaded}
\begin{Highlighting}[]
\NormalTok{i}\OtherTok{\textless{}{-}}\DecValTok{1} \CommentTok{\#Set an index number to 1 for the first subject}
\NormalTok{y.frame}\OtherTok{\textless{}{-}}\NormalTok{dat[i,}\DecValTok{5}\SpecialCharTok{:}\DecValTok{11}\NormalTok{,drop}\OtherTok{=}\ConstantTok{FALSE}\NormalTok{] }\CommentTok{\#ith row, columns 5 through 11}
\NormalTok{y.frame }\CommentTok{\#Check indifference points}
\end{Highlighting}
\end{Shaded}

\begin{verbatim}
##       y1     y7    y30    y90   y365  y1825  y9125
## 1 0.4922 0.9922 0.9454 0.8984 0.8984 0.3984 0.1016
\end{verbatim}

\begin{Shaded}
\begin{Highlighting}[]
\NormalTok{y}\OtherTok{\textless{}{-}}\FunctionTok{as.vector}\NormalTok{(}\FunctionTok{as.matrix}\NormalTok{(}\FunctionTok{t}\NormalTok{(y.frame))) }\CommentTok{\#Make y a vector}
\NormalTok{D}\OtherTok{\textless{}{-}}\FunctionTok{c}\NormalTok{(}\DecValTok{1}\NormalTok{,}\DecValTok{7}\NormalTok{,}\DecValTok{30}\NormalTok{,}\DecValTok{90}\NormalTok{,}\DecValTok{365}\NormalTok{,}\DecValTok{1825}\NormalTok{,}\DecValTok{9125}\NormalTok{) }\CommentTok{\#Define the delays D}
\FunctionTok{plot}\NormalTok{(D,y, }\CommentTok{\#scatter plot. First argument is horizontal axis, second is vertical axis}
     \AttributeTok{xlab=}\StringTok{\textquotesingle{}Delay (days)\textquotesingle{}}\NormalTok{,}\AttributeTok{ylab=}\StringTok{\textquotesingle{}Indifference point\textquotesingle{}}\NormalTok{) }\CommentTok{\# Label axes}
\end{Highlighting}
\end{Shaded}

\includegraphics{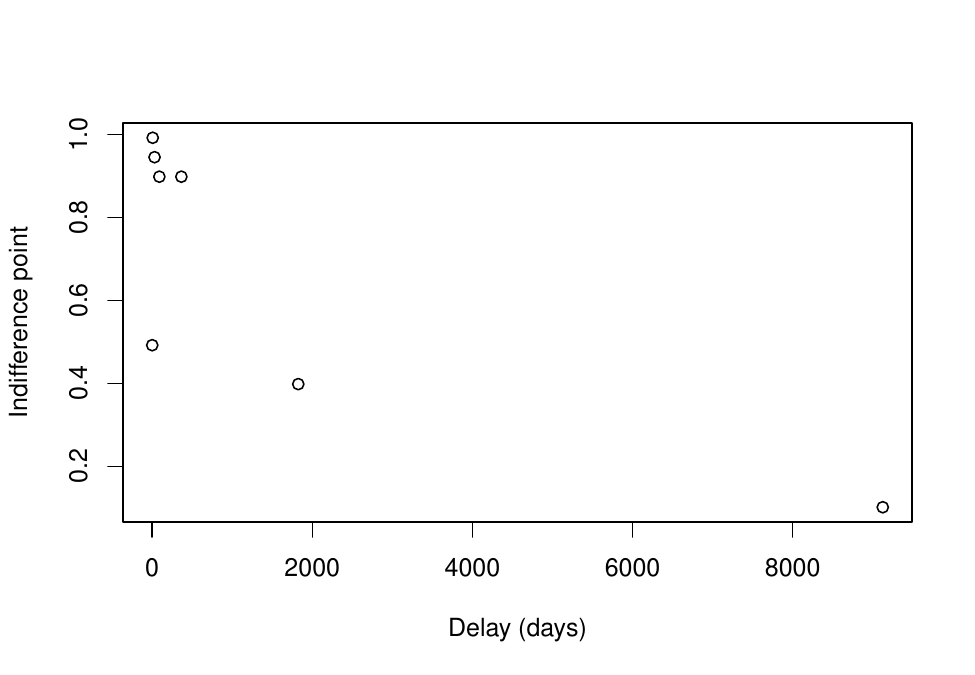}

Our plot for the indifference points and delays of the first participant
reveals that generally, as delay increases, indifference point
decreases. There is a notable jump between one day and one week delay,
however. One might expect that indifference points would get lower and
lower as delay increases (reflecting the reduced utility a larger later
reward has at longer delays). In practice, not all observed discounting
data follow this trend. Unusual or unexpected discounting patterns might
emerge due to lack of attention on the part of the participant, a single
mistaken response early in a sequence of a titrating questionnaire, the
participant having failed to understand the task's directions, or due to
other considerations a participant has that the experimenter cannot be
aware of (e.g.~an unexpectedly high indifference point might be due to
an upcoming event like a medical bill the participant is thinking
about.)

\emph{Visualizing discounting on the log-delay scale}

If you are anything like me, your first algebra teacher tried to tell
you how important the logarithm function is. You ignored them since you
figured you'd never use this abstract information on a daily basis. If
you are reading this sentence, then the day has arrived when you will be
using the logarithm function on a daily basis.

The logarithmic function (which is the inverse function of the
exponential function) is central to modern data analytic practices. It
simplifies mathematical optimization, computationally stabilizes numbers
extremely close to zero, and aids in visualization for many types of
analyses in many fields, including visualizing delay discounting data by
log transforming delay. In the present case, we have a troublesome
vertical jump among the indifference points early in the delay sequence
that is incredibly difficult to appreciate visually in the scatter plot
above. Let us consider the same scatter plot but put the natural log of
delay on the horizontal axis.

\begin{Shaded}
\begin{Highlighting}[]
\FunctionTok{plot}\NormalTok{(}\FunctionTok{log}\NormalTok{(D),y, }
     \AttributeTok{xlab=}\StringTok{\textquotesingle{}ln(Delay)\textquotesingle{}}\NormalTok{,}\AttributeTok{ylab=}\StringTok{\textquotesingle{}Indifference point\textquotesingle{}}\NormalTok{) }\CommentTok{\# Label axes}
\end{Highlighting}
\end{Shaded}

\includegraphics{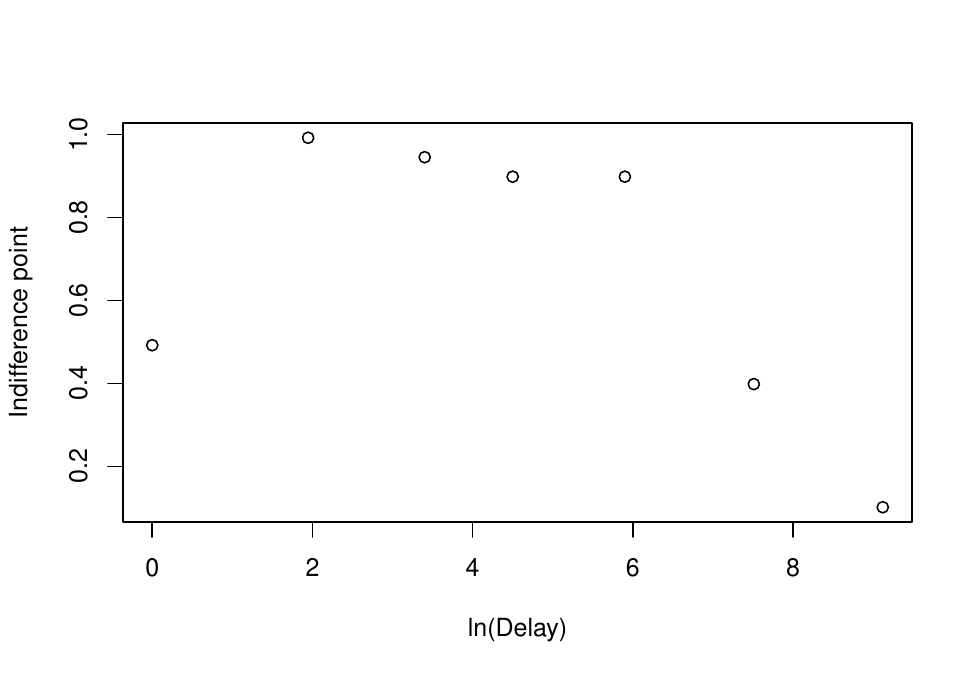}

(Brief note: In R syntax, ``log()'' is the natural log function
(frequently denoted ``ln()'' mathematically), while the ``log10()''
function is log base 10 in R.)

The logarithm function spreads out numbers that are closer to zero, and
compresses numbers that are further from zero. Thus, very short
distances between delays early on (such as one day and one week) get
spread out while more remote distances (e.g.~five years and 25 years to
the right side of the panel) are compressed when the log is taken. Thus,
the reader can see that on the natural log scale, delays are more
equally spaced and we can assess patterns of indifference points among
the shorter delays with much greater ease. (Further information and
exercises to review the logarithm function can be found at the end of
this chapter.)

Specifically, we can now easily see that the first participant's jump in
indifference point between a day and a week is very large, indicating
the apparent contradiction. This participant reports quickly devaluing a
reward for a delay of 1 day but exhibiting essentially no discounting of
a reward in a week. This is not the idealized behavior we expect for
delay discounting, and this jump violates one of the Johnson \& Bickel
criteria (Johnson and Bickel 2008).

The purpose of the Johnson \& Bickel criteria is to give researchers a
simple method to identify patterns of indifference points that may be
inconsistent with expected behavior of delay discounting. Quoting that
paper, the Johnson \& Bickel criteria would flag a data set if:
``\ldots either or both of the two following criterion were met: 1) if
any indifference point (starting with the second delay) was greater than
the preceding indifference point by a magnitude greater than 20\% of the
larger later reward\ldots'' and ``2) if the last (i.e., 25-year)
indifference point was not less than the first (1 day or 1 week,
depending on the study) indifference point by at least a magnitude equal
to 10\% of the larger later reward\ldots''

As with any other data-analytic exercise focusing on unusual data
points, outliers, etc. the decision about how to proceed with analysis
in the face of unexpected, unusual, or atypical data is ultimately
subjective, but should be considered prior to conducting analysis. Going
forward, we will analyze the full set of data without excluding
participants who violate of Johnson \& Bickel criteria or attention
checks. This is meant to illustrate overall analyses without being
overly prescriptive about how to handle non-systematic data. Note that
there are exercises at the end of the chapter that guide the motivated
reader to re-conduct the presented analyses with various exclusion
criteria in place and to determine to what extent changing exclusion
criteria can alter the analysis conclusions. Inclusion criteria are set
prior to analysis in a typical study and not second-guessed midstream.
In research practice, one does not simply change things at will and try
every combination of analyses until a preconceived conclusion is
reached. This is called ``p-hacking,'' and more information can be found
in the \textbf{Contemporary issues in statistical practice} Section
below.

We shouldn't automatically delete any data which is unusual, because it
is still valid if collected according to the research protocols.
However, be mindful about situations in which a small number of data
points yield an out-sized influence on study conclusions is also not
ideal. The point of statistical analysis is to aggregate information
from many subjects to make broader conclusions about the population as a
whole, and if a few individual data points change or obscure our view of
the entire picture, have we really focused on the broader population or
just a handful of unusual points? Perhaps unsurprisingly, decisions
about how to handle participant data that does not follow expected
discounting patterns can impact the final conclusions of a study.

We next turn our attention to fitting discounting models. Several models
have been proposed. We first illustrate the widely used hyperbolic
discounting model, sometimes referred to as the ``Mazur model'' (Mazur
1987), which is

\[E(y)=\frac{A}{1 + k*D},\]

where \(E(y)\) is the expected value of the indifference point \(y\),
i.e., the regression line value at delay \(D\). The value \(A\) is the
amount of the delayed reward, and \(k\) is the discounting rate. Note
that the hyperbolic model fits the data in terms of un-logged delay, not
\(\text{ln(Delay)}\) as was previously visualized. If \(k=0\) then the
participant does not devalue the reward as a function of delay. The data
we analyze in this chapter has \(A\) already, so our indifference points
are between zero and one and \(A\) is replaced by \(1\) subsequently.

The hyperbolic discounting function is just one example of a functional
form that describes delay discounting. Exercise (2) at the end of this
chapter explores another popular discounting function. For more
information on other discounting functions, see (McKerchar et al. 2009).

When interpreting the output of data analyses and drawing inferences to
populations larger than the sample that was collected, it is vitally
important to clearly distinguish between unknown population parameters
and sample statistics that estimate those parameters. The parameter
\(k\) is unknown and must be estimated on the basis of delay data \(D\)
and indifference point data \(y\). We follow the convention in most
statistical texts by denoting data-based estimates of unknown parameters
with a hat. Thus, we use the symbol \(\hat{k}\) to denote the data-based
statistics that estimates the true (but unknown) parameter \(k\). We
follow this convention for parameters throughout this chapter.

To begin to make some of the above statistical concepts more concrete,
we now fit the Mazur model to the first participants' data using
nonlinear regression via the ``nls'' (which stands for ``nonlinear least
squares'') function. We will fit the model, plot the fit alongside
indifference points, and extract the estimated discounting rate
\(\hat{k}\).

\begin{Shaded}
\begin{Highlighting}[]
\NormalTok{mod}\OtherTok{\textless{}{-}}\FunctionTok{nls}\NormalTok{(y}\SpecialCharTok{\textasciitilde{}}\DecValTok{1}\SpecialCharTok{/}\NormalTok{(}\DecValTok{1}\SpecialCharTok{+}\NormalTok{k}\SpecialCharTok{*}\NormalTok{D),}\AttributeTok{start=}\FunctionTok{list}\NormalTok{(}\AttributeTok{k=}\NormalTok{.}\DecValTok{1}\NormalTok{)) }\CommentTok{\#Fit the Mazur model to these data}
\NormalTok{mod }\CommentTok{\#View the estimated value of k}
\end{Highlighting}
\end{Shaded}

\begin{verbatim}
## Nonlinear regression model
##   model: y ~ 1/(1 + k * D)
##    data: parent.frame()
##         k 
## 0.0007053 
##  residual sum-of-squares: 0.2733
## 
## Number of iterations to convergence: 8 
## Achieved convergence tolerance: 2.116e-06
\end{verbatim}

\begin{Shaded}
\begin{Highlighting}[]
\CommentTok{\#Verify computation of residual sum{-}of{-}squares}
\FunctionTok{sum}\NormalTok{(((y}\SpecialCharTok{{-}}\FunctionTok{predict}\NormalTok{(mod))}\SpecialCharTok{\^{}}\DecValTok{2}\NormalTok{))}
\end{Highlighting}
\end{Shaded}

\begin{verbatim}
## [1] 0.2732825
\end{verbatim}

The above code creates an object called ``mod'' that stores the results
of a nonlinear regression. The R syntax \(y\sim 1/(1+k*D)\) indicates
that we want to fit indifference points \(y\) as the outcome variable
with the model form following the Mazur model. By running the object
``mod'', we see a reminder of the form of the model we chose (useful
when several models are being considered), an estimate
\(\hat{k}=0.0007053\), residual sum of squares (which quantify the sum
of squared residuals, i.e.~the vertical distance between each point and
a regression line squared and added up), and some information about
model convergence. We say more about least squares and the model fitting
that is happening ``under-the-hood'' in a few paragraphs.

Speaking candidly, I have always had difficulty intuitively
understanding the discount rate \(k\). For this participant, the
estimate \(\hat{k}=0.0007\). The discounting rate \(k\) is higher among
individuals who discount rapidly, and lower among individuals who do
not. Beyond that, \(k\) is not particularly interpretable for most
people. A related metric, called the \emph{Effective Delay 50} (ED50)
describes the length of delay for which the participant would forfeit
half of the larger later reward to have the reward immediately (Yoon and
Higgins 2008). Conveniently, \(\text{ED50}=1/k\) when the Mazur model is
in use. Thus, a data-based estimator for \(\text{ED50}\) is
\(\hat{\text{ED50}}=1/\hat{k}\). Effective delay 50 is important due to
its interpretability, the fact that its interpretation is the same among
competing discounting models (Franck et al. 2015), and for its role and
implementation in the development of a very brief but effective
discounting questionnaire (Koffarnus and Bickel 2014) that titrates to
ED50 then infers \(\hat{k}\) based on the form of Mazur's model rather
than by obtaining indifference points directly.

Now let's have a look at the line of best fit for these data and also
visualize and compute ED50. For longer code chunks like this one, feel
free to run the code one line at a time to learn how each command works.

\begin{Shaded}
\begin{Highlighting}[]
\NormalTok{D.s}\OtherTok{\textless{}{-}}\FunctionTok{seq}\NormalTok{(}\DecValTok{0}\NormalTok{,}\DecValTok{9500}\NormalTok{,}\DecValTok{1}\NormalTok{) }\CommentTok{\#Create a fine grid across delays. Used later to plot the regression line.}
\NormalTok{preds}\OtherTok{\textless{}{-}}\FunctionTok{predict}\NormalTok{(mod,}\AttributeTok{newdata=}\FunctionTok{data.frame}\NormalTok{(}\AttributeTok{D=}\NormalTok{D.s)) }\CommentTok{\#Obtain predicted values for each grid point}
\FunctionTok{plot}\NormalTok{(D,y,}\AttributeTok{xlab=}\StringTok{\textquotesingle{}Delay (days)\textquotesingle{}}\NormalTok{,}\AttributeTok{ylab=}\StringTok{\textquotesingle{}Indifference point\textquotesingle{}}\NormalTok{,}
     \AttributeTok{main=}\StringTok{"Indifference points, model fit, and ED50"}\NormalTok{) }\CommentTok{\#Scatter plot}
\FunctionTok{lines}\NormalTok{(D.s,preds,}\AttributeTok{col=}\StringTok{\textquotesingle{}red\textquotesingle{}}\NormalTok{) }\CommentTok{\#Add the regression line to the plot}

\DocumentationTok{\#\#Add a legend}
\FunctionTok{legend}\NormalTok{(}\AttributeTok{x=}\DecValTok{4500}\NormalTok{,}\AttributeTok{y=}\NormalTok{.}\DecValTok{8}\NormalTok{,}\AttributeTok{legend=}\FunctionTok{c}\NormalTok{(}\StringTok{"Indifference point"}\NormalTok{,}\StringTok{"Regression line"}\NormalTok{),}\AttributeTok{col=}\FunctionTok{c}\NormalTok{(}\StringTok{"black"}\NormalTok{,}\StringTok{"red"}\NormalTok{),}
       \AttributeTok{pch=}\FunctionTok{c}\NormalTok{(}\DecValTok{1}\NormalTok{,}\ConstantTok{NA}\NormalTok{),}\AttributeTok{lty=}\FunctionTok{c}\NormalTok{(}\ConstantTok{NA}\NormalTok{,}\DecValTok{1}\NormalTok{)) }



\NormalTok{k.hat}\OtherTok{\textless{}{-}}\FunctionTok{summary}\NormalTok{(mod)}\SpecialCharTok{$}\NormalTok{coef[}\DecValTok{1}\NormalTok{,}\DecValTok{1}\NormalTok{] }\CommentTok{\#Store k.hat}
\NormalTok{ED50}\OtherTok{=}\DecValTok{1}\SpecialCharTok{/}\NormalTok{k.hat }\CommentTok{\#Store ED50}
\NormalTok{k.hat}
\end{Highlighting}
\end{Shaded}

\begin{verbatim}
## [1] 0.0007052959
\end{verbatim}

\begin{Shaded}
\begin{Highlighting}[]
\NormalTok{ED50}
\end{Highlighting}
\end{Shaded}

\begin{verbatim}
## [1] 1417.845
\end{verbatim}

\begin{Shaded}
\begin{Highlighting}[]
\CommentTok{\#Add lines and text to illustrate ED50}
\FunctionTok{lines}\NormalTok{(}\AttributeTok{x=}\FunctionTok{c}\NormalTok{(ED50,ED50),}\AttributeTok{y=}\FunctionTok{c}\NormalTok{(}\DecValTok{0}\NormalTok{,}\FloatTok{0.5}\NormalTok{),}\AttributeTok{lty=}\DecValTok{3}\NormalTok{)}
\FunctionTok{lines}\NormalTok{(}\AttributeTok{x=}\FunctionTok{c}\NormalTok{(}\SpecialCharTok{{-}}\DecValTok{2000}\NormalTok{,ED50),}\AttributeTok{y=}\FunctionTok{c}\NormalTok{(}\FloatTok{0.5}\NormalTok{,}\FloatTok{0.5}\NormalTok{),}\AttributeTok{lty=}\DecValTok{3}\NormalTok{)}
\FunctionTok{text}\NormalTok{(}\DecValTok{2700}\NormalTok{,}\FloatTok{0.54}\NormalTok{,}\FunctionTok{paste}\NormalTok{(}\StringTok{\textquotesingle{}ED50 =\textquotesingle{}}\NormalTok{,}\FunctionTok{round}\NormalTok{(ED50),}\StringTok{\textquotesingle{}days\textquotesingle{}}\NormalTok{))}
\end{Highlighting}
\end{Shaded}

\includegraphics{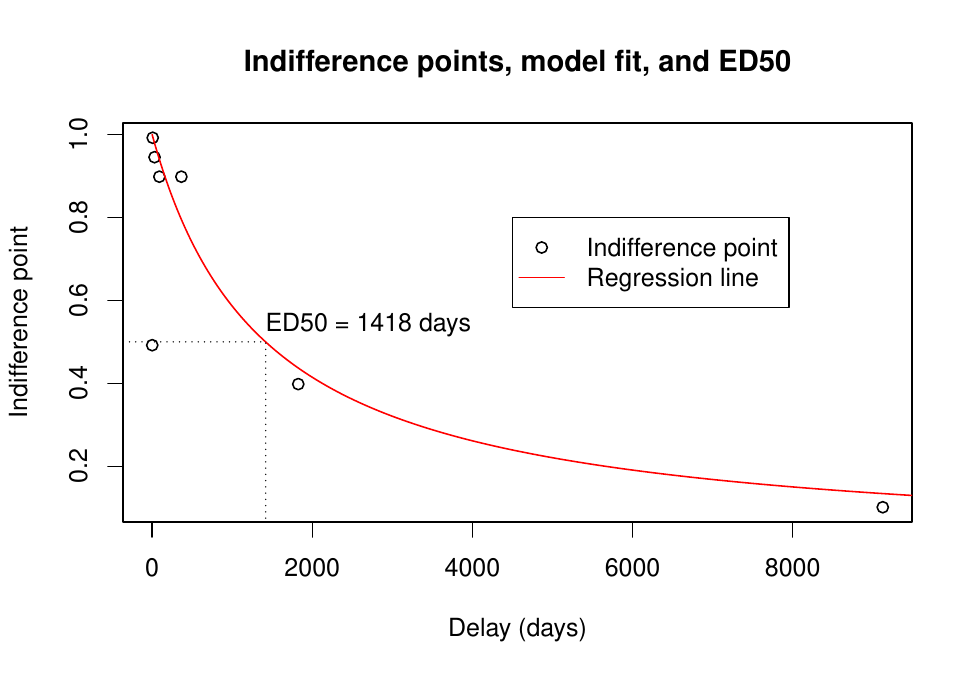}

We estimate \(\hat{\text{ED50}}=\frac{1}{\hat{k}}=1418\) days. The plot
includes the indifference points, the nonlinear regression line, and a
visual depiction of the ED50 value which corresponds to the point on the
regression line where indifference point is 0.5.

Before we proceed with similar analyses for the remaining study
participants' data, we describe a few more statistical concepts that are
under the hood. The above plot cogently shows that the goal is to
determine which value of \(k\) provides the ``best fitting line'' to the
data. For basic analyses, this is frequently accomplished by
\textbf{least squares}. We will focus initially on explaining this
approach. Other methods to estimate parameters include \textbf{maximum
likelihood estimation} and \textbf{Bayesian} methods, which have been
described in the context of discounting here (Franck et al. 2023) and
here (Franck et al. 2019), respectively.

In order to place a regression line ``close'' to observed data points,
we must have some notion of collective distance between the set of data
points and the line. Our nonlinear \textbf{least squares} approach uses
\textbf{residual sum-of-squares} (RSS) as a notion of distance. RSS is
calculated by (i) taking the difference between each data point and the
corresponding point on the regression line with the same delay \(D\),
(2) squaring those individual differences (so everything is positive and
points below the line do not `cancel out' points above the line), and
finally (3) adding up these squared differences.

Mathematically,

\[\text{RSS}=\sum_{i=1}^n(y_i-\hat{y}_i)^2\] where \(n\) is the sample
size, \(y_i\) is the \(ith\) indifference point, \(\hat{y}_i\) is the
value of the regression line corresponding to the \(ith\) indifference
point. The large sample operator \(\sum_{i=1}^n\) indicates summation of
all squared differences for the data set.

RSS describes how far a given regression line is from the observed data
in a least squares sense. In order to find the least squares line, one
must search the space of all possible parameter values (\(k\) in this
problem) in order to find the value \(\hat{k}\) that produces a line
with the smallest possible RSS. The field of mathematics that studies
strategies to determine optimal values of criterion as a function of the
criterion's inputs is called \textbf{optimization}. The good news is
that (i) least squares approaches are mathematically easy to implement
for the Mazur model, and (ii) we have already shown how to use R to
obtain optimal \(\hat{k}\) values for discounting data (using the
``nls'' function above).

For those who would like more concrete demonstration that we indeed have
the best possible line and who would also like to see some more R code,
see the code below.

\begin{Shaded}
\begin{Highlighting}[]
\CommentTok{\#define a function that computes RSS on basis of provided k for observed data}
\NormalTok{RSS}\OtherTok{\textless{}{-}}\ControlFlowTok{function}\NormalTok{(k)\{ }
\NormalTok{  yhat}\OtherTok{\textless{}{-}}\DecValTok{1}\SpecialCharTok{/}\NormalTok{(}\DecValTok{1}\SpecialCharTok{+}\NormalTok{k}\SpecialCharTok{*}\NormalTok{D)}
\NormalTok{  resid}\OtherTok{\textless{}{-}}\NormalTok{y}\SpecialCharTok{{-}}\NormalTok{yhat}
\NormalTok{  rss}\OtherTok{\textless{}{-}}\FunctionTok{sum}\NormalTok{((y}\SpecialCharTok{{-}}\NormalTok{yhat)}\SpecialCharTok{\^{}}\DecValTok{2}\NormalTok{)}
  \FunctionTok{return}\NormalTok{(rss)}
\NormalTok{\}}

\NormalTok{k.seq}\OtherTok{\textless{}{-}}\FunctionTok{seq}\NormalTok{(}\DecValTok{0}\NormalTok{,.}\DecValTok{01}\NormalTok{,.}\DecValTok{0001}\NormalTok{) }\CommentTok{\#sequence of k values}
\NormalTok{RSS.seq}\OtherTok{\textless{}{-}}\FunctionTok{sapply}\NormalTok{(k.seq,RSS) }\CommentTok{\#sequence of RSS values at each k in sequence}
\FunctionTok{plot}\NormalTok{(k.seq,RSS.seq,}\AttributeTok{type=}\StringTok{\textquotesingle{}l\textquotesingle{}}\NormalTok{,}\AttributeTok{ylab=}\StringTok{\textquotesingle{}RSS\textquotesingle{}}\NormalTok{,}\AttributeTok{xlab=}\StringTok{\textquotesingle{}k\textquotesingle{}}\NormalTok{,}
     \AttributeTok{main=}\StringTok{"Lowest RSS occurs at estimated k value"}\NormalTok{)}
\FunctionTok{abline}\NormalTok{(}\AttributeTok{v=}\NormalTok{k.hat,}\AttributeTok{col=}\StringTok{\textquotesingle{}red\textquotesingle{}}\NormalTok{,}\AttributeTok{lty=}\DecValTok{3}\NormalTok{)}
\FunctionTok{optimize}\NormalTok{(RSS,}\AttributeTok{interval=}\FunctionTok{c}\NormalTok{(}\DecValTok{0}\NormalTok{,}\DecValTok{100}\NormalTok{))}
\end{Highlighting}
\end{Shaded}

\begin{verbatim}
## $minimum
## [1] 0.0007296753
## 
## $objective
## [1] 0.2734153
\end{verbatim}

\begin{Shaded}
\begin{Highlighting}[]
\CommentTok{\#add a legend}
\FunctionTok{legend}\NormalTok{(}\AttributeTok{x=}\NormalTok{.}\DecValTok{002}\NormalTok{,}\AttributeTok{y=}\NormalTok{.}\DecValTok{2}\NormalTok{,}\AttributeTok{legend=}\FunctionTok{c}\NormalTok{(}\StringTok{"RSS as a function of k"}\NormalTok{, }\StringTok{"k.hat from previous analysis"}\NormalTok{),}
       \AttributeTok{col=}\FunctionTok{c}\NormalTok{(}\StringTok{"black"}\NormalTok{,}\StringTok{"red"}\NormalTok{),}\AttributeTok{lty=}\FunctionTok{c}\NormalTok{(}\DecValTok{1}\NormalTok{,}\DecValTok{3}\NormalTok{))}
\end{Highlighting}
\end{Shaded}

\includegraphics{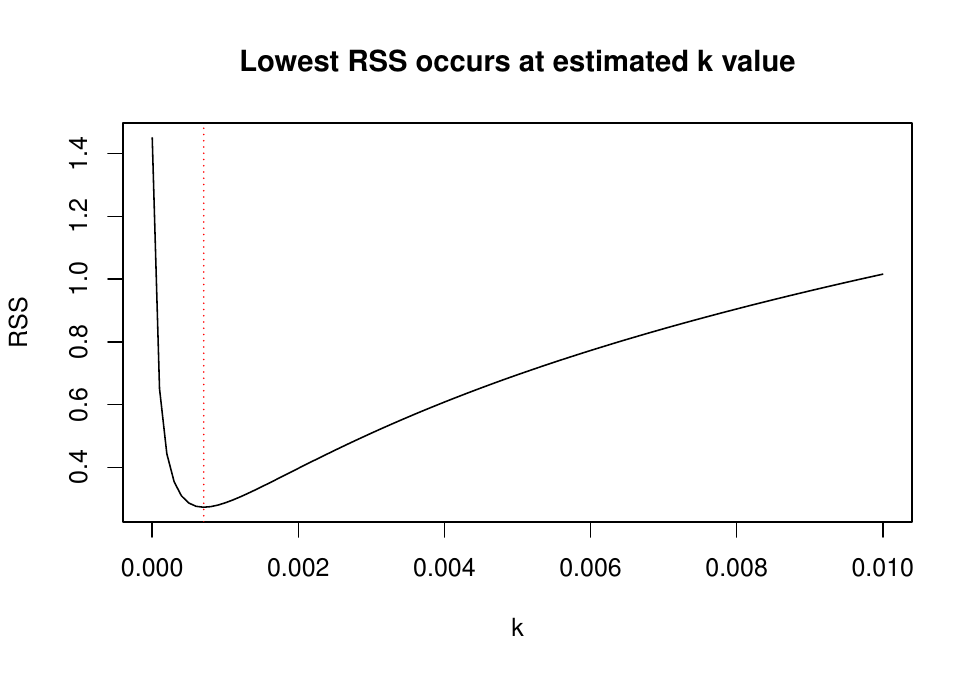}

\emph{Area under the curve as a measure of discounting}

Thus far, we have quantified discounting using the value \(k\) from the
Mazur equation and ED50. Another approach which does not assume a
specific functional form involves computing area under the discounting
curve (AUC) (Myerson, Green, and Warusawitharana 2001). The idea is to
join up the indifference points with line segments, then calculate the
area of the region below this curve. Using the data from subject 1:

\begin{Shaded}
\begin{Highlighting}[]
\FunctionTok{plot}\NormalTok{(D,y, }\CommentTok{\#scatter plot. First argument is horizontal axis, second is vertical axis}
     \AttributeTok{xlab=}\StringTok{\textquotesingle{}Delay (days)\textquotesingle{}}\NormalTok{,}\AttributeTok{ylab=}\StringTok{\textquotesingle{}Indifference point\textquotesingle{}}\NormalTok{) }\CommentTok{\# Label axes}
\FunctionTok{lines}\NormalTok{(D,y)}
\FunctionTok{lines}\NormalTok{(D,y,}\AttributeTok{type=}\StringTok{\textquotesingle{}h\textquotesingle{}}\NormalTok{,}\AttributeTok{lty=}\DecValTok{2}\NormalTok{,}\AttributeTok{col=}\StringTok{\textquotesingle{}lightgray\textquotesingle{}}\NormalTok{)}
\end{Highlighting}
\end{Shaded}

\includegraphics{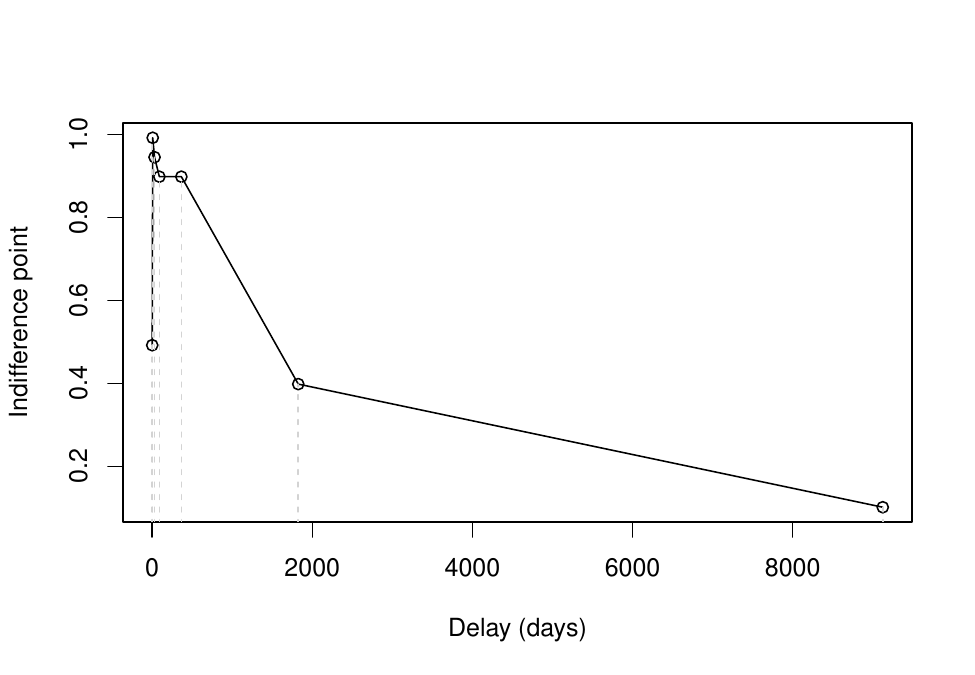} It
is clear from the above picture that the region under the adjoined line
segments consists of adjoined trapezoids. Computing the area of
trapezoids and adding them up is easy, which is why this strategy for
computing area under connect-the-dots style curves is called the
\emph{trapezoidal rule.} (Note that in calculus, the trapezoidal rule
produces close approximations to the area under smooth functions by
evaluating the function on a dense grid, connecting the dots, and adding
up the area of the trapezoids). This can be conveniently obtained in R
using the \emph{trapz} function in the \emph{pracma} package.

While base R comes with many capabilities, there are also a great number
of add-on libraries that can be accessed free of charge that further
extend R's capabilities. Many available packages are stored on the
Comprehensive R Archive Network (CRAN). If you are connected to the
internet, you can access the CRAN repository by clicking ``Tools
--\textgreater{} Install Packages\ldots{}'' menu through RStudio, then
using the menu options to choose among available packages.
Alternatively, one can use the
\texttt{install.packages(“package\_name”)} command in the console to
access available packages. If you have a goal in mind but do not know
the name of an R package that meets that goal, then it is usually a good
idea to do some Google searching to determine whether a suitable package
is available. In the present case, the \emph{pracma} package should be
obtained and installed before running the following code.

\begin{Shaded}
\begin{Highlighting}[]
\FunctionTok{library}\NormalTok{(pracma)}
\FunctionTok{trapz}\NormalTok{(D,y)}
\end{Highlighting}
\end{Shaded}

\begin{verbatim}
## [1] 3100.774
\end{verbatim}

The AUC for these data is is 3100.774.

AUC has some pros and some cons. For pros, it is easy to compute. AUC is
directly associated with discounting since more devaluation of rewards
across delays leads to shorter trapezoids which in turn leads to lower
AUC.

AUC is unlike a typical statistical regression approach, because AUC
does not attempt to describe an underlying function and quantify the
departure of observed outcome data (indifference points in this case)
from that function. Thus, the AUC metric is not tied to any specific
theoretical framework (to quote the authors' abstract). Whether this
non-theoretical take on discounting is a pro or con, the availability of
AUC is at minimum a useful empirical benchmark to compare other
theoretical models' quantification of discounting. There is a practice
problem at the end of the chapter that tasks the reader with evaluating
the association among various discounting metrics including AUC.

In terms of cons, AUC alone does not enforce or even evaluate whether
rewards are devalued as a function of delay. A participant whose
indifference points are increasing could end up with an AUC value very
similar to a participant who discounts sensibly. A valid straight-line
pattern of indifference points decreasing from top left to bottom right
would have the same AUC as an invalid increasing line from bottom left
to top right, for example.

Another downside of AUC is that, as a non-theoretical approach, it is
difficult to definitively resolve best practices on theoretical grounds.
For example, an analyst may be concerned that the shorter delays form
trapezoids that are much smaller than the longer delays and thus
relatively short delays may be severely under weighted when computing
AUC. This provides useful information in the delay range where rewards
are being devalued rapidly. But this also produces very narrow trapezoid
bases, thus arguably the most important delays play a much smaller role
in AUC than the subsequent delays which are spaced out more. A natural
approach to increase the emphasis on early delays might be to compute an
area under the curve metric on the basis of logged delays (Borges et al.
2016), as presented in the code below.

\begin{Shaded}
\begin{Highlighting}[]
\FunctionTok{plot}\NormalTok{(}\FunctionTok{log}\NormalTok{(D),y, }\CommentTok{\#scatter plot. First argument is horizontal axis, second is vertical axis}
     \AttributeTok{xlab=}\StringTok{\textquotesingle{}ln(Delay)\textquotesingle{}}\NormalTok{,}\AttributeTok{ylab=}\StringTok{\textquotesingle{}Indifference point\textquotesingle{}}\NormalTok{) }\CommentTok{\# Label axes}
\FunctionTok{lines}\NormalTok{(}\FunctionTok{log}\NormalTok{(D),y)}
\FunctionTok{lines}\NormalTok{(}\FunctionTok{log}\NormalTok{(D),y,}\AttributeTok{type=}\StringTok{\textquotesingle{}h\textquotesingle{}}\NormalTok{,}\AttributeTok{lty=}\DecValTok{2}\NormalTok{,}\AttributeTok{col=}\StringTok{\textquotesingle{}lightgray\textquotesingle{}}\NormalTok{)}
\end{Highlighting}
\end{Shaded}

\includegraphics{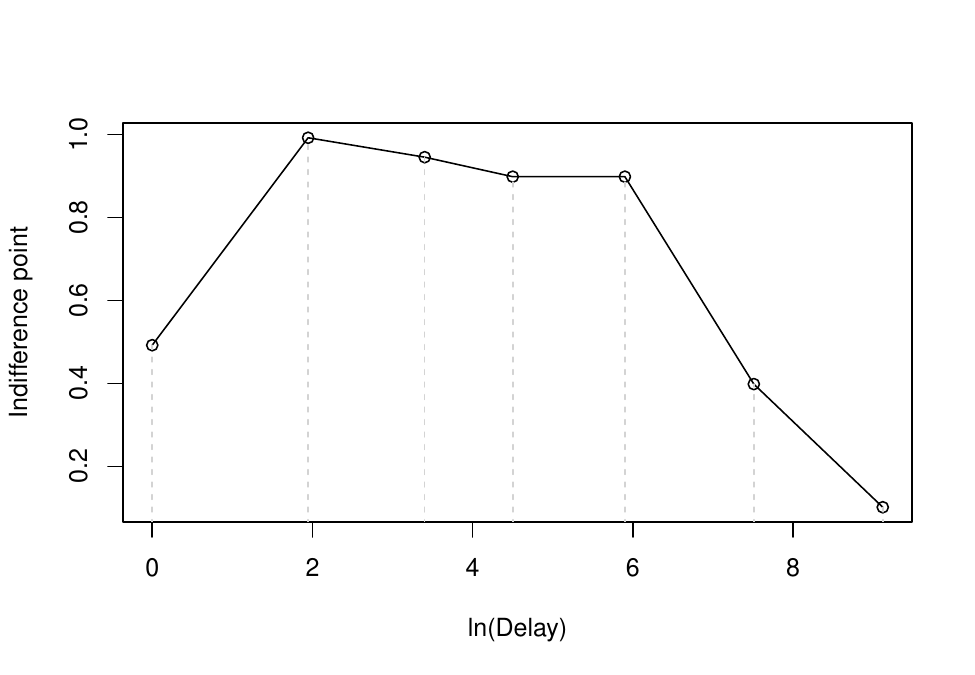}

\begin{Shaded}
\begin{Highlighting}[]
\FunctionTok{trapz}\NormalTok{(}\FunctionTok{log}\NormalTok{(D),y)}
\end{Highlighting}
\end{Shaded}

\begin{verbatim}
## [1] 6.570705
\end{verbatim}

The analyst who computed AUC on the basis of unlogged delays got
3100.774 in unlogged delay space. The analyst who was concerned this
gave too small weight to the short delays got 6.57 but in logged delay
space. Who is right and who is wrong? I have no idea, and since there is
no underlying theory for AUC, decisions such as these have no definitive
resolution. We can just argue about best practices.

Another potential downside is the inability for AUC to quantify error
variance in the Stage 1 fit. A hallmark of statistical reasoning is the
notion that observed data are a noisy realization arising from some
underlying data-generating model. Deliberately characterizing observed
data with a model with an unrealistically low estimate of error variance
(zero in fact) might be viewed as inelegant.

\textbf{Don't we need to complete the above analyses for the remaining
105 participants' data?}

Stage 1 has been focused on obtaining discounting metrics for a
participant based on their indifference point data. Stage 2 will analyze
these metrics as a function of other predictors (e.g., determine whether
smokers discount faster than non-smokers on average). Thus, we must
automate the Stage 1 analyses so we can quickly obtain discounting
metrics for all participants in the sample before we move on to Stage 2.

Rather than copy-pasting similar commands to those above for each of the
106 participants' data, we will instead embed our code in a \textbf{for
loop}. A for loop is a general programming technique that allows the
user to repeat a set of user-specified commands for a pre-specified
number of iterations. We will see that once the analysis plan is in
place for a single subject, repeating that same analysis a great number
of times is actually easy.

The for loops typically used in analyzing discounting data typically
involve defining an index value that increments at each iteration of the
loop. We use the letter \emph{i}. The index \emph{i} is typically set to
the value one initially, then it ticks up each iteration. Commands are
written to operate on, e.g., the \emph{ith} individual's data (currently
the first participant), results are stored for a later-on analysis, the
value of \emph{i} increments to 2, the same commands are applied to the
second participant's data, and so on until each participant's data are
analyzed and results stored.

Below find the syntax to make a for loop that captures all participants'
estimated \(\hat{k}\) values. We create an empty vector called
``K.vec''. The shorthand ``vec'' stands for ``vector''. In R, a vector
is a list of either numeric or character variables where the order of
the entries is important. In our code, K.vec is initially empty. The for
loop will store the first participant's estimated \(\hat{k}\) value in
the first position of K.vec, the second participant's \(\hat{k}\) goes
in the second position of K.vec, and so on.

\begin{Shaded}
\begin{Highlighting}[]
\NormalTok{K.vec}\OtherTok{\textless{}{-}}\FunctionTok{c}\NormalTok{() }\CommentTok{\#initialize K.vec as an empty vector}
\ControlFlowTok{for}\NormalTok{(i }\ControlFlowTok{in} \DecValTok{1}\SpecialCharTok{:}\DecValTok{106}\NormalTok{)\{ }\CommentTok{\#begin at i=1, do everything in curly brackets, increment i, repeat}
\NormalTok{  y.frame}\OtherTok{\textless{}{-}}\NormalTok{dat[i,}\DecValTok{5}\SpecialCharTok{:}\DecValTok{11}\NormalTok{,drop}\OtherTok{=}\ConstantTok{FALSE}\NormalTok{] }\CommentTok{\#pull indifference points for ith participant}
\NormalTok{  y}\OtherTok{\textless{}{-}}\FunctionTok{as.vector}\NormalTok{(}\FunctionTok{as.matrix}\NormalTok{(}\FunctionTok{t}\NormalTok{(y.frame)))  }\CommentTok{\#turn data frame y.frame into vector y}
\NormalTok{  K.vec[i]}\OtherTok{\textless{}{-}}\FunctionTok{summary}\NormalTok{(}\FunctionTok{nls}\NormalTok{(y}\SpecialCharTok{\textasciitilde{}}\NormalTok{(}\DecValTok{1}\SpecialCharTok{+}\NormalTok{K}\SpecialCharTok{*}\NormalTok{D)}\SpecialCharTok{\^{}}\NormalTok{(}\SpecialCharTok{{-}}\DecValTok{1}\NormalTok{),}\AttributeTok{start=}\FunctionTok{list}\NormalTok{(}\AttributeTok{K=}\NormalTok{.}\DecValTok{1}\NormalTok{)))}\SpecialCharTok{$}\NormalTok{coef[}\DecValTok{1}\NormalTok{,}\DecValTok{1}\NormalTok{] }\CommentTok{\#fit Mazur model}
                                                                    \CommentTok{\#store k estimate}
\NormalTok{\}}
\end{Highlighting}
\end{Shaded}

With each participant's value of \(\hat{k}\) stored in the corresponding
element of the vector K.vec, let's examine these values and confirm that
there are indeed 106 of them stored.

\begin{Shaded}
\begin{Highlighting}[]
\CommentTok{\#print the k.hat values inside K.vec to the console}
\NormalTok{K.vec}
\end{Highlighting}
\end{Shaded}

\begin{verbatim}
##   [1] 7.052959e-04 1.952525e-02 6.279889e-03 3.106512e+00 9.443459e-04
##   [6] 1.183173e-02 7.792406e-03 4.121380e-02 2.166866e-03 1.851430e-03
##  [11] 1.835216e-02 6.848412e-02 1.253200e-02 5.644911e-05 3.917017e-03
##  [16] 2.750789e-02 9.416390e-03 2.706837e-03 1.158562e-01 2.339159e-04
##  [21] 5.654420e-02 1.924157e-02 7.786736e-04 2.225344e-03 1.238148e-02
##  [26] 1.737184e-03 1.006645e-02 6.653795e-02 4.119001e-04 4.115331e-04
##  [31] 2.122284e-02 8.613770e-02 1.438784e+00 2.457380e-02 1.087262e-03
##  [36] 4.645629e-03 4.105892e-02 9.843393e-03 2.102845e-03 3.947859e-01
##  [41] 4.392440e-03 1.043863e-02 5.207565e-03 2.480166e-03 2.890474e-01
##  [46] 7.969068e-03 1.041180e-06 1.111135e-02 1.334191e-03 4.690139e-02
##  [51] 1.541998e-02 2.656261e-03 2.676365e-03 1.582058e-01 2.288003e-03
##  [56] 1.606234e-02 1.598051e-02 5.757535e-02 1.524680e-02 5.524457e-01
##  [61] 1.029969e-03 4.620361e-01 1.040897e-02 3.163147e-04 5.727909e-03
##  [66] 1.660060e-01 1.531857e-02 5.343197e-01 3.719899e-02 1.558544e-01
##  [71] 1.345135e-03 3.329768e-04 4.723152e-05 3.592244e-02 8.398586e-01
##  [76] 3.267308e+00 1.283517e-03 2.957418e-02 5.499067e-03 1.423051e-01
##  [81] 2.190507e-03 3.898910e-03 6.109505e-04 5.744756e-04 1.192669e-02
##  [86] 3.503430e-04 5.695165e-02 1.393875e-02 1.192544e-01 1.404184e-04
##  [91] 4.431371e-04 1.813248e-02 3.385442e-03 4.639105e-03 1.151348e-02
##  [96] 1.204211e-02 1.318007e-03 5.991004e-03 1.141583e-02 6.315233e-04
## [101] 1.683690e-02 4.873263e-04 4.684651e-03 1.041180e-06 1.500254e-03
## [106] 6.310517e-03
\end{verbatim}

\begin{Shaded}
\begin{Highlighting}[]
\CommentTok{\#confirm there are n=106 observations in K.vec}
\FunctionTok{length}\NormalTok{(K.vec)}
\end{Highlighting}
\end{Shaded}

\begin{verbatim}
## [1] 106
\end{verbatim}

\begin{Shaded}
\begin{Highlighting}[]
\CommentTok{\#Add the k.hat values to the data set}
\NormalTok{dat}\SpecialCharTok{$}\NormalTok{k}\OtherTok{\textless{}{-}}\NormalTok{K.vec}
\end{Highlighting}
\end{Shaded}

R uses a specific format for scientific notation to express numbers that
are very small or very large. For example,
\(\text{7.05e-04}=7.05\times10^{-4}=0.000705\).

Once the above code has been run, the loop is complete and the vector
``K.vec'' is a length 106 vector where each element corresponds to that
participant number's \(\hat{k}\) value. Now that \(\hat{k}\) values have
been obtained for all participant data we move on the the second stage
of analysis.

\emph{Probability discounting}

We have focused most of our attention on the analysis of delay
discounting data, where we quantify the rate of devaluation of a reward
as a function of delay to that reward. \emph{Probability discounting}
instead quantifies the rate of devaluation of a reward as a function of
the probability of not receiving the reward. For example, in (Rachlin,
Raineri, and Cross 1991), study participants were asked to choose
between hypothetical smaller but certain cash rewards and larger
uncertain rewards. For example, \$500 for sure or a 50\% chance of
receiving \$1,000. Such prompts vary the probability of winning, express
this probability in terms of odds against, and use discounting functions
to quantify discounting rates. As with delay discounting, a
one-parameter hyperbolic equation has been proposed. Also, like delay
discounting, several models for probability discounting have been
proposed. The extent to which probability and delay discounting are
comparable phenomena is also a topic of discussion in the literature.
See (Killeen 2023) and the references therein for more discussion.

\textbf{Stage 2}

In Stage 2, we analyze the collection of estimated discounting rates
(the \(\hat{k}\) values) for the participants, including as a function
of potential predictors. These estimated discounting rates are
considered as data for the second stage of analysis. For this
illustration we will compare discounting rates (i) between males and
females, (ii) between smokers and non-smokers, and (iii) as a function
of age. With these goals in mind, we begin with an \textbf{exploratory
data analysis}. Exploratory data analysis typically begins by plotting
data. We consider both univariate plots and also multivariate plots.

A good first rule for exploratory data analysis is to establish which
scale of measurement the data are measured on. There are frequently many
potential variables that could be collected with respect to a research
question. For example, in these data, smoking status was measured as a
binary yes/no variable. However, researchers could instead ask how many
cigarettes participants smoke in a given week. This latter question
would be measured using a numeric variable that would be a whole number
greater than or equal to zero. Analyzing data appropriately on the scale
they are measured is a core tenet of a properly conducted analysis. In
this study, we have variables that were measured on a binary scale
(smoking status and gender) and two variables measures on a numeric
scale (age in years and discounting rate \(\hat{k}\)).

\emph{Univariate analyses}

We'll begin our overview of exploratory data analysis by illustrating
\textbf{univariate} summaries and graphics. Univariate approaches focus
on understanding each variables' data alone without any assessment of
association with other variables. We may be interested in, for example,
determining the number of participants who are smokers, how many are
male and female, what the distribution of observed ages are, and also
the distribution of observed \(\hat{k}\) values.

We may wish to tabulate the number of participants in each category of
our categorical variables.

Note that in R, the dollar sign notation allows the user to specify an
object inside another object. In this case, ``dat\$gender'' reads the
``gender'' variable out of the ``dat'' object.

\begin{Shaded}
\begin{Highlighting}[]
\FunctionTok{table}\NormalTok{(dat}\SpecialCharTok{$}\NormalTok{gender) }\CommentTok{\#make a table of gender data from the dat object}
\end{Highlighting}
\end{Shaded}

\begin{verbatim}
## 
## Female   Male 
##     27     79
\end{verbatim}

\begin{Shaded}
\begin{Highlighting}[]
\FunctionTok{table}\NormalTok{(dat}\SpecialCharTok{$}\NormalTok{smoke\_cigs) }\CommentTok{\#make a table of smoking status from the dat object}
\end{Highlighting}
\end{Shaded}

\begin{verbatim}
## 
##  No Yes 
##  52  54
\end{verbatim}

There are 27 females and 79 males in this sample. There are 54 smokers
and 52 nonsmokers in this sample.

Now let us consider the distribution of age. \textbf{Histograms} are a
useful graphic to show features of a distribution of numeric data.
Histograms organize the range of observed data into bins, then a bar is
constructed for each bin to reflect the number of participants in that
bin. We can also compute summary statistics for age.

\begin{Shaded}
\begin{Highlighting}[]
\FunctionTok{hist}\NormalTok{(dat}\SpecialCharTok{$}\NormalTok{age,}\AttributeTok{main=}\StringTok{"Histogram of age"}\NormalTok{,}\AttributeTok{xlab=}\StringTok{"Age"}\NormalTok{)}
\end{Highlighting}
\end{Shaded}

\includegraphics{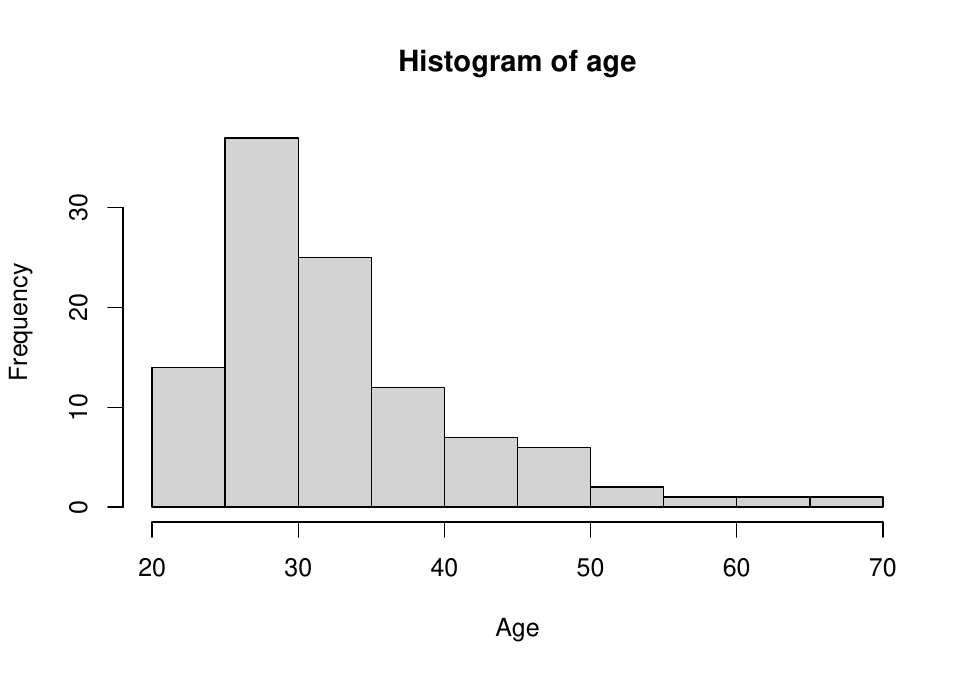}

\begin{Shaded}
\begin{Highlighting}[]
\FunctionTok{summary}\NormalTok{(dat}\SpecialCharTok{$}\NormalTok{age)}
\end{Highlighting}
\end{Shaded}

\begin{verbatim}
##    Min. 1st Qu.  Median    Mean 3rd Qu.    Max. 
##   21.00   28.00   31.00   33.49   36.00   67.00
\end{verbatim}

We can see that ages range from 21 to 67 years with a mean age of 33.49
years, etc. The histogram reveals that the distribution of age is not
symmetric. It has a heavier right tail, i.e., it is skewed to the right.
The bars can be interpreted to indicate (for example) that about 14
individuals have an age between 20 and 25.

Let us next consider the distribution of \(\hat{k}\) obtained in Stage
1.

\begin{Shaded}
\begin{Highlighting}[]
\FunctionTok{hist}\NormalTok{(K.vec,}\AttributeTok{main=}\StringTok{"Estimated k values from n=106 participants"}\NormalTok{,}
     \AttributeTok{xlab=}\StringTok{\textquotesingle{}k\textquotesingle{}}\NormalTok{,}\AttributeTok{breaks=}\DecValTok{20}\NormalTok{) }\CommentTok{\#make a histogram }
\end{Highlighting}
\end{Shaded}

\includegraphics{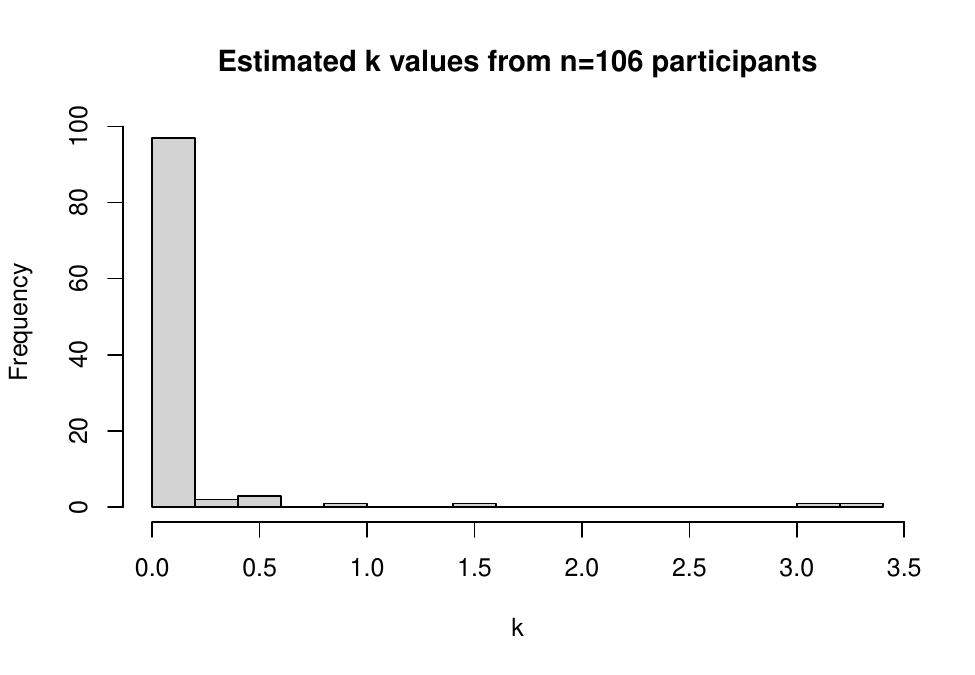}

The above histogram shows the 106 estimated k values. It is visually
apparent that the distribution of these \(\hat{k}\) values is skewed
right and bounded below by zero. For these reasons, it is customary to
analyze k values on the natural log scale. The R code below also
superimposes a normal distribution density function on to the
\(\hat{k}\) values. This density curve is centered at the sample mean
with sample standard deviation also obtained from the data.

\begin{Shaded}
\begin{Highlighting}[]
\FunctionTok{hist}\NormalTok{(}\FunctionTok{log}\NormalTok{(K.vec),}\AttributeTok{main=}\StringTok{"Estimated ln(k) values from the n=106 participants"}\NormalTok{,}
     \AttributeTok{xlab=}\StringTok{\textquotesingle{}ln(k)\textquotesingle{}}\NormalTok{,}\AttributeTok{breaks=}\DecValTok{20}\NormalTok{,}\AttributeTok{freq=}\ConstantTok{FALSE}\NormalTok{)}
\NormalTok{xbar}\OtherTok{\textless{}{-}}\FunctionTok{mean}\NormalTok{ (}\FunctionTok{log}\NormalTok{(K.vec));s}\OtherTok{\textless{}{-}}\FunctionTok{sqrt}\NormalTok{(}\FunctionTok{var}\NormalTok{(}\FunctionTok{log}\NormalTok{(K.vec)))}
\NormalTok{ks}\OtherTok{\textless{}{-}}\FunctionTok{seq}\NormalTok{(}\SpecialCharTok{{-}}\DecValTok{15}\NormalTok{,}\DecValTok{5}\NormalTok{,.}\DecValTok{1}\NormalTok{)}
\FunctionTok{lines}\NormalTok{(ks,}\FunctionTok{dnorm}\NormalTok{(ks,}\AttributeTok{mean=}\NormalTok{xbar,s),}\AttributeTok{col=}\StringTok{\textquotesingle{}red\textquotesingle{}}\NormalTok{)}
\end{Highlighting}
\end{Shaded}

\includegraphics{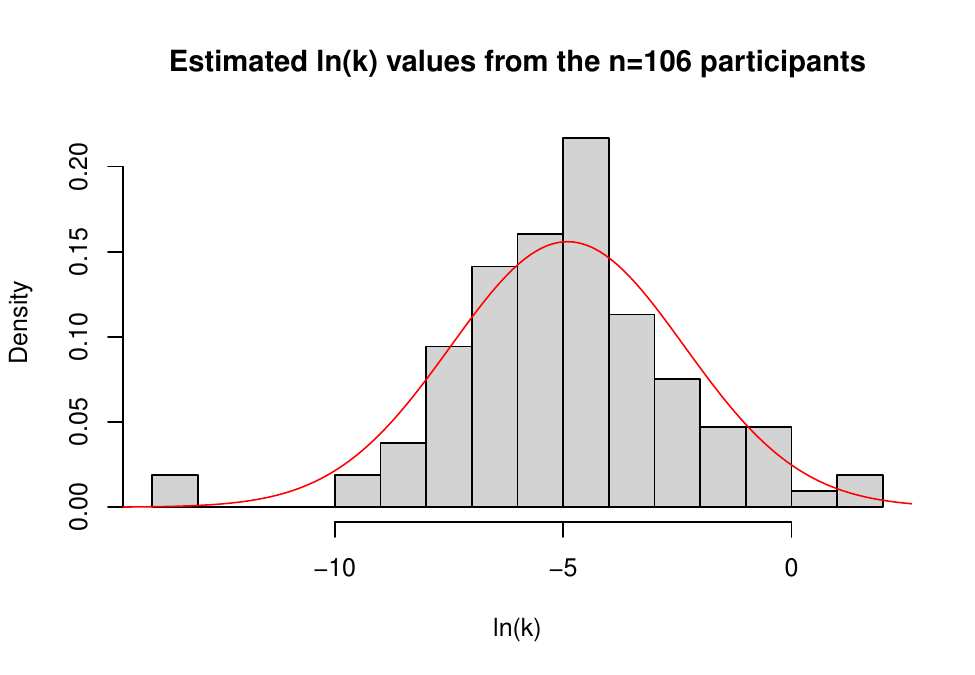}

The above histogram of \(\text{ln}(k)\) is more symmetric and
bell-shaped than the original version, and thus more suitable for
typical statistical analyses which assume data are normally distributed.
Analytic approaches that assume normality include include t-tests and
corresponding confidence intervals, analysis of variance (ANOVA). and
typical regression approaches. These rudimentary analyses are presented
in most intro textbooks, but rest assured more sophisticated statistical
techniques have been developed and continue to be developed to model
data of every sort no matter the distribution.

A subtle but very important statistical point: the appearance of a
bell-shaped curve is typically assumes for \textbf{residuals} (i.e., the
difference between a data point and an average or regression line)
rather than the raw data themselves. To elaborate, the normality
assumption requires slightly more than for us to look for a plausible
normal distribution in a univariate histogram. For two-sample t-tests
and ANOVA-based approaches, we presume the outcome variable
\(\text{ln}(\hat{k})\) is distributed normally within each group.
Ordinary regression problems model the outcome variable
\(\text{ln}(\hat{k})\) as a function of (potentially several) predictor
variables, and the extent to which individual data points depart from
the line (i.e., residuals) are assumed to be normally distributed with
constant variance. We will explore these techniques in more depth
subsequently, including an assessment of whether the appropriate
normality conditions hold in an end-of-chapter exercise. It is generally
true that \(\text{ln}(\hat{k})\) usually satisfies normality assumptions
adequately and certainly more than the \(\hat{k}\) values, and so we can
proceed with these sorts of techniques and models.

Many statistical techniques exist for settings where data are not
normally distributed, but many of the most well-known techniques do
assume a normal distribution. For the sake of brevity, we will not
consider alternative techniques in depth. These include non-parametric
rank-based procedures, quantile regression, and other techniques
specifically developed to handle non-normal data, such as logistic
regression for binary outcomes. Simple two group comparisons, such as
the rank-based Mann-Whitney test are available that do not assume
normally distributed data. However, rank-based tests do not scale up
well for multiple predictors. In these data, a researcher may be
interested in modeling delay discounting as a function of smoking
status, gender, and age simultaneously. Rank-based inference is not
readily available, even for a small problem like this.

\emph{Multivariate Analyses}

Examining the association among variables is central to scientific and
statistical practice. While the above univariate analyses helped
familiarize us with the data and anticipate potential challenges with
subsequent analysis (e.g., needing to log transform the \(\hat{k}\)
data), we are centrally interested in the association among the
variables. To proceed, let us consider each pairwise association in
these data. With four variables, there are six pairwise associations.

As with univariate associations, the scales of measurement for each
variable imply which graphical approaches may be sensible. When both
variables are numeric, a scatter plot is typically a good choice. When
there is one categorical variable and one numeric variable, we might
consider producing a box plot. When both variables are categorical,
mosaic plots are a viable choice. There are a variety of other viable
choices as well, but these three graphics are fundamental.

\begin{Shaded}
\begin{Highlighting}[]
\NormalTok{lnk}\OtherTok{\textless{}{-}}\FunctionTok{log}\NormalTok{(dat}\SpecialCharTok{$}\NormalTok{k)}
\FunctionTok{plot}\NormalTok{(dat}\SpecialCharTok{$}\NormalTok{age,lnk,}\AttributeTok{xlab=}\StringTok{"Age"}\NormalTok{)}
\end{Highlighting}
\end{Shaded}

\includegraphics{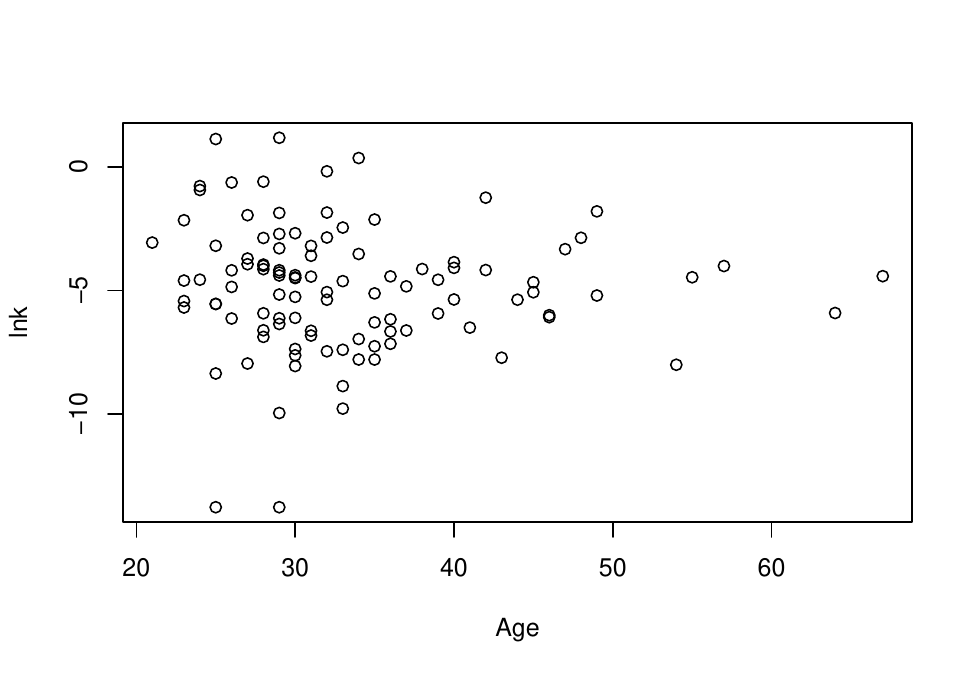}

\begin{Shaded}
\begin{Highlighting}[]
\FunctionTok{cor}\NormalTok{(dat}\SpecialCharTok{$}\NormalTok{age,lnk)}
\end{Highlighting}
\end{Shaded}

\begin{verbatim}
## [1] -0.04820073
\end{verbatim}

The scatter plot does not appear to show a particularly strong
relationship between age and \(\text{ln}(\hat{k})\). The cor() function
computes the \textbf{correlation coefficient} between two variables,
which measures the strength and direction of the linear relationship and
is always between -1 and 1. Negative value show an inverse relationship
and positive values show a direct relationship between the variables.
The closer further the correlation is from zero, the stronger the
relationship. Here, we see the correlation is \(r=-0.048\), indicating a
slight negative association in these data.

\begin{Shaded}
\begin{Highlighting}[]
\FunctionTok{par}\NormalTok{(}\AttributeTok{mfrow=}\FunctionTok{c}\NormalTok{(}\DecValTok{2}\NormalTok{,}\DecValTok{2}\NormalTok{)) }\CommentTok{\#Make a 2X2 figure}
\FunctionTok{boxplot}\NormalTok{(dat}\SpecialCharTok{$}\NormalTok{age}\SpecialCharTok{\textasciitilde{}}\NormalTok{dat}\SpecialCharTok{$}\NormalTok{gender,}\AttributeTok{xlab=}\StringTok{"Gender"}\NormalTok{,}\AttributeTok{ylab=}\StringTok{"Age"}\NormalTok{)}
\FunctionTok{boxplot}\NormalTok{(dat}\SpecialCharTok{$}\NormalTok{age}\SpecialCharTok{\textasciitilde{}}\NormalTok{dat}\SpecialCharTok{$}\NormalTok{smoke\_cigs,}\AttributeTok{xlab=}\StringTok{"Smoker"}\NormalTok{,}\AttributeTok{ylab=}\StringTok{"Age"}\NormalTok{)}

\FunctionTok{boxplot}\NormalTok{(lnk}\SpecialCharTok{\textasciitilde{}}\NormalTok{dat}\SpecialCharTok{$}\NormalTok{gender,}\AttributeTok{xlab=}\StringTok{"Gender"}\NormalTok{)}
\FunctionTok{boxplot}\NormalTok{(lnk}\SpecialCharTok{\textasciitilde{}}\NormalTok{dat}\SpecialCharTok{$}\NormalTok{smoke\_cigs,}\AttributeTok{xlab=}\StringTok{"Smoker"}\NormalTok{)}
\end{Highlighting}
\end{Shaded}

\includegraphics{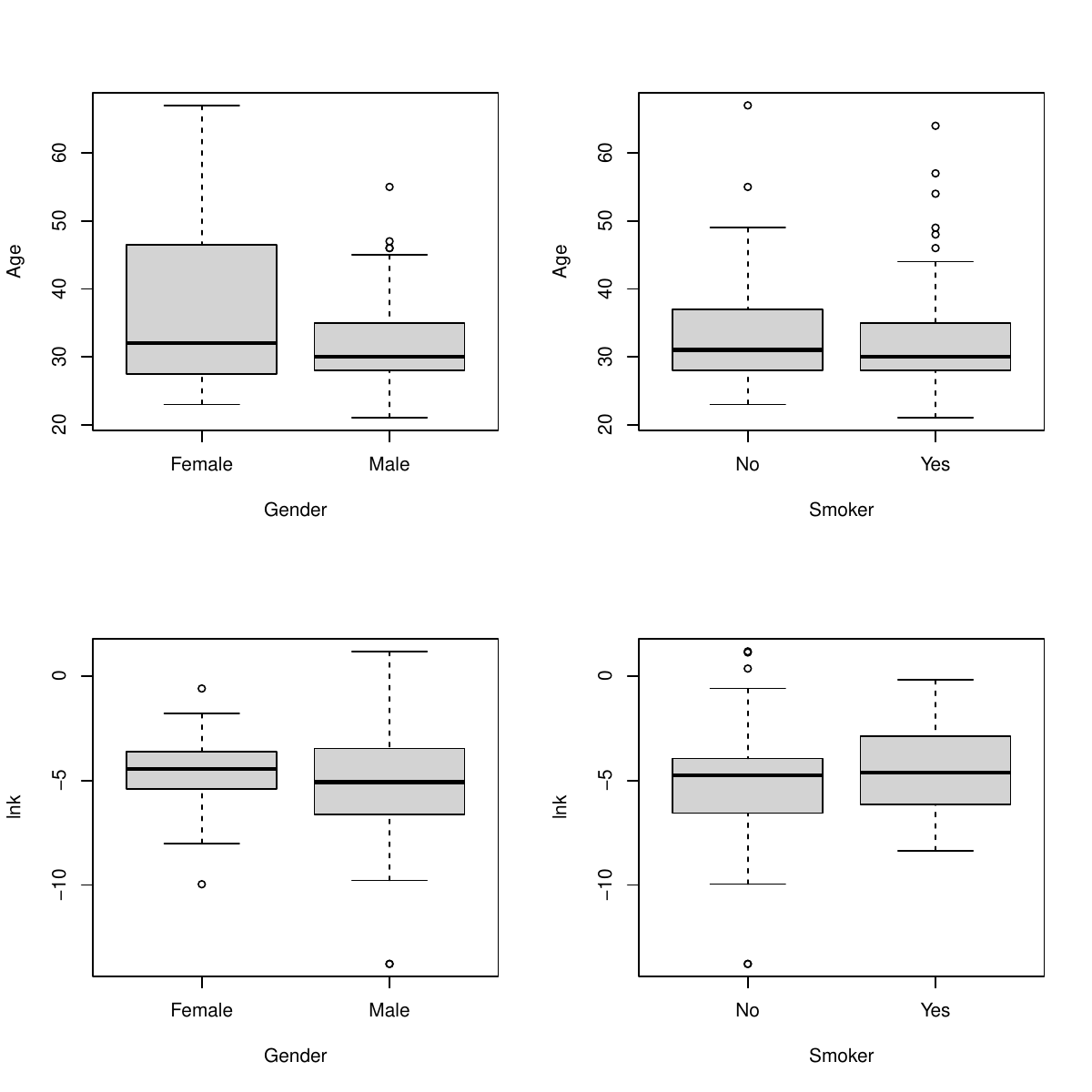}

The age box plots indicate that females tended to be older than males in
this study, but there was no substantive difference in age between
smokers and non-smokers. The \(\text{ln}(\hat{k})\) box plots show a
very slight difference in logged discounting rates, with females in this
study slightly higher than males, and smokers slightly higher than
non-smokers.

Mosaic plots can be a useful graphic when multiple categorical
variables' are being considered. A mosaic plot consists of a series of
rectangular regions organized to show the relative size of all
combinations of the levels of the categorical variables (female smokers,
female non-smokers, male smokers, and male non-smokers in this case).
The code below shows this. An upside of the mosaic plot is that it is
easy to visually interpret the areas of rectangles compared for example
with the pie-slice regions of a pie chart.

\begin{Shaded}
\begin{Highlighting}[]
\NormalTok{tab}\OtherTok{\textless{}{-}}\FunctionTok{table}\NormalTok{(dat}\SpecialCharTok{$}\NormalTok{gender,dat}\SpecialCharTok{$}\NormalTok{smoke\_cigs)}
\FunctionTok{mosaicplot}\NormalTok{(tab,}\AttributeTok{main=}\StringTok{\textquotesingle{}Mosaic plot\textquotesingle{}}\NormalTok{,}\AttributeTok{xlab=}\StringTok{\textquotesingle{}Gender\textquotesingle{}}\NormalTok{,}\AttributeTok{ylab=}\StringTok{\textquotesingle{}Smoker\textquotesingle{}}\NormalTok{)}
\end{Highlighting}
\end{Shaded}

\includegraphics{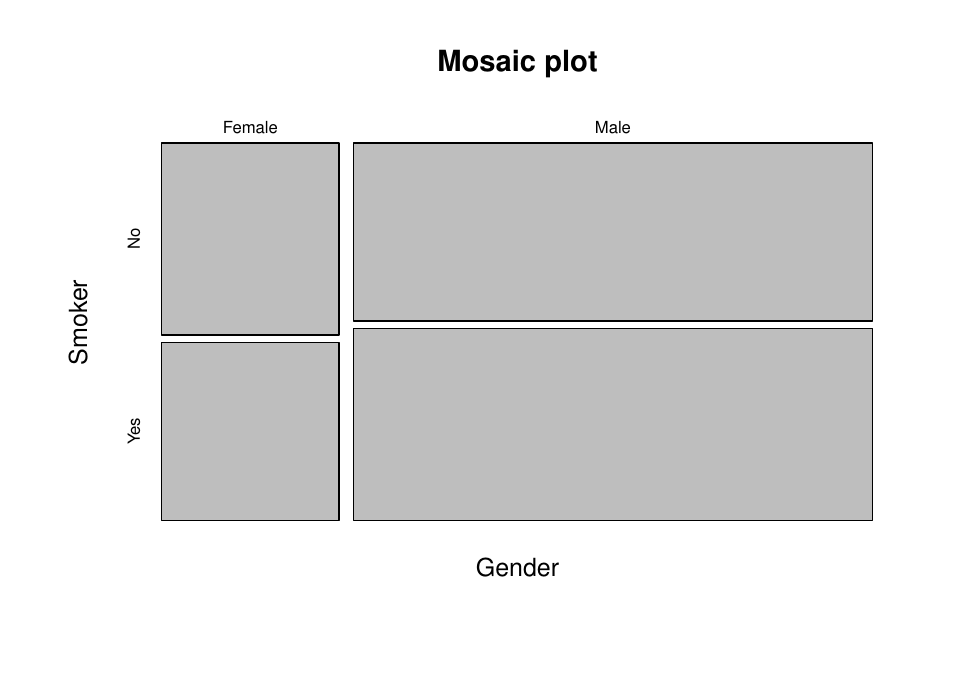}

The above mosaic plot re-expresses that smokers and non-smokers are
roughly evenly split in these data, that there are more males than
females. The gender split among smokers resembles that among non-smokers
(and the overall split), and similarly, the smoking status split is
similar for both genders, suggesting that these variables do not share
much of an association. If, for example, almost all of the smokers were
male and the non-smokers were female, this would be reflected in the
plot with very large rectangles in one diagonal, small rectangles in the
other, and would indicate a strong association.

\emph{A note about plots}

For our demonstration, we will load the ``ggplot2'' package and the
``GGally'' package. Both of these packages provide the user with
sophisticated graphical options. The analysis code we are adapting for
this example and related discussion appears here:
(\url{https://ggobi.github.io/ggally/reference/ggpairs.html}). Once
those have been acquired and installed as described above, the following
code will enable their use within R.

\begin{Shaded}
\begin{Highlighting}[]
\FunctionTok{library}\NormalTok{(ggplot2)}
\FunctionTok{library}\NormalTok{(GGally)}
\end{Highlighting}
\end{Shaded}

\begin{verbatim}
## Registered S3 method overwritten by 'GGally':
##   method from   
##   +.gg   ggplot2
\end{verbatim}

The function we will use is called ``ggpairs''. To read the help file
documentation, run the following code, first removing the comment
symbol.

\begin{Shaded}
\begin{Highlighting}[]
\CommentTok{\#?ggpairs}
\end{Highlighting}
\end{Shaded}

Next we pass a data frame that includes age, gender, smoker, and
\(\text{ln}(\hat{k})\) to the ``ggpairs'' function

\begin{Shaded}
\begin{Highlighting}[]
\NormalTok{plot.frame}\OtherTok{\textless{}{-}}\FunctionTok{data.frame}\NormalTok{(}\AttributeTok{age=}\NormalTok{dat}\SpecialCharTok{$}\NormalTok{age,}\AttributeTok{gender=}\NormalTok{dat}\SpecialCharTok{$}\NormalTok{gender,}\AttributeTok{smoker=}\NormalTok{dat}\SpecialCharTok{$}\NormalTok{smoke\_cigs,}\AttributeTok{lnk=}\FunctionTok{log}\NormalTok{(dat}\SpecialCharTok{$}\NormalTok{k))}

\NormalTok{plot.ob }\OtherTok{\textless{}{-}} \FunctionTok{ggpairs}\NormalTok{(}
\NormalTok{  plot.frame,}
  \AttributeTok{upper =} \FunctionTok{list}\NormalTok{(}\AttributeTok{continuous =} \StringTok{"density"}\NormalTok{, }\AttributeTok{combo =} \StringTok{"box\_no\_facet"}\NormalTok{),}
  \AttributeTok{lower =} \FunctionTok{list}\NormalTok{(}\AttributeTok{continuous =} \StringTok{"points"}\NormalTok{, }\AttributeTok{combo =} \StringTok{"dot\_no\_facet"}\NormalTok{)}
\NormalTok{)}
\NormalTok{plot.ob}
\end{Highlighting}
\end{Shaded}

\includegraphics{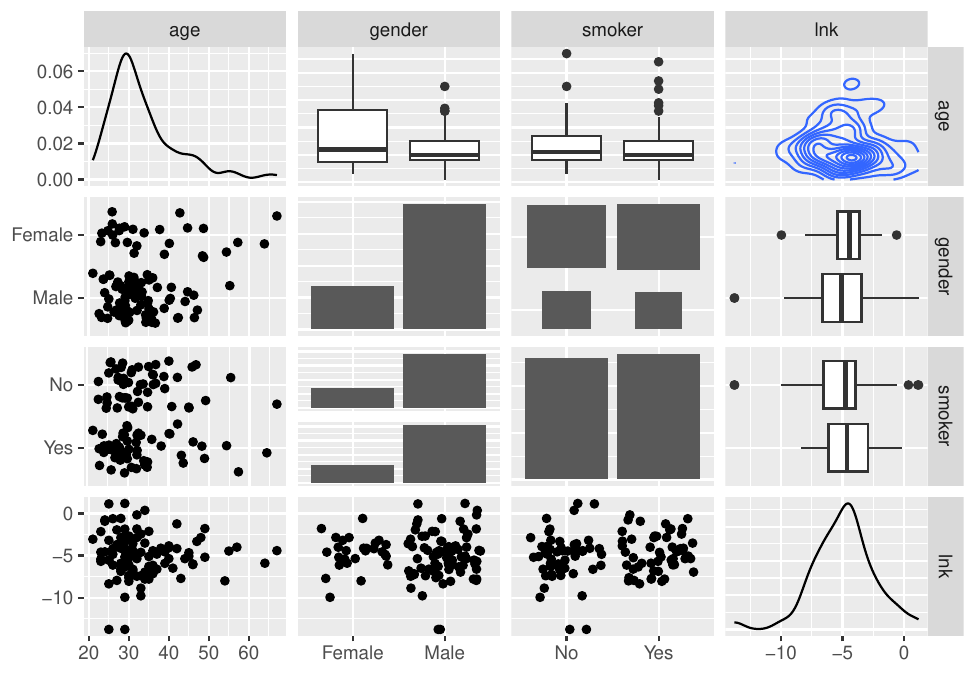}

The above plot summarizes all two-way associations and uses a variety of
graphical plotting techniques. It is rife with visual information. This
is a sophisticated example of a \textbf{pairs plot}. In a pairs plot,
variables (age, gender, smoking status, \(\text{ln}(\hat{k})\) in this
case) indicates the position a variable is summarized in both rows and
columns. The diagonal panels from top left to bottom right are
univariate summaries of each variable. All other panels are bivariate
associations, where the reader looks ``left-to-right'' to learn what is
plotted on the vertical axis and ``up-and-down'' to determine what is on
the horizontal axis.

Starting from the top left and moving across rows, the first panel is a
density plot of age. The second and third panels in the first row are
box plots of gender by age and smoking status by age, respectively. The
fourth panel in the first row is a contour plot of
\(\text{ln}(\hat{k})\) by age.

The first panel in the second row is a jitter plot showing individual
data points for age by gender. A small amount of random noise (i.e.,
jitter) is added to points on the vertical axis here to help unstack
points within gender groups. The second row second column panel is a bar
plot of gender. Next is a mosaic plot of smoking status by gender, then
an \(\text{ln}(\hat{k})\) by gender box plot.

The third row includes a jitter plot of age by smoking status, bar plots
of gender by smoking status, bar plots of smoking status, a bar plot of
smoking status, and a box plot of \(\text{ln}(\hat{k})\) by smoking
status.

The fourth row include a scatter plot of age by \(\text{ln}(\hat{k})\),
jitter plots of gender by \(\text{ln}(\hat{k})\) and smoking status by
\(\text{ln}(\hat{k})\), and a univariate density plot of
\(\text{ln}(\hat{k})\).

Creating useful and beautiful graphics is both a science and an art. The
initial plots we made convey core data analytic insight but are not
especially visually appealing and are arguably not a judicious use of
space for most publication venues. (We commonly agonize over attempts to
perfect plots for publication.) By contrast, the ``GGally'' pairs plot
looks sleeker (with many additional plotting options available - see
documentation), but is potentially dense to the point of being hard to
digest. Details such as axis labels for the inner bar plots must be
minimized or omitted in order to fit other information. When making
plots we recommend organizing the core plots to emphasize and add
clarity to the core points of the writing, and relegate other
potentially useful plots (e.g., the pairs plot above) to an appendix to
satisfy extra-curious readers. For the purpose of teaching the basics,
we have opted to include code for fairly simple plots in this chapter.
These plots adequately convey data-analytic insight but are probably not
appealing enough to be publication quality in general.

\textbf{Descriptive versus inferential statistical approaches}

Broadly speaking, statistical statements are \emph{Descriptive} or
\emph{Inferential}. \textbf{Descriptive statistics} describe sample data
only.

\begin{Shaded}
\begin{Highlighting}[]
\FunctionTok{mean}\NormalTok{(lnk)}
\end{Highlighting}
\end{Shaded}

\begin{verbatim}
## [1] -4.904101
\end{verbatim}

In the above example code, we compute the sample mean
\(\text{ln}(\hat{k})=-4.90\). A \emph{descriptive} interpretation of
this number is ``Among the 106 participants in our research study, the
average \(\text{ln}(\hat{k})\)'' value is -4.90.'' We simply summarize
what we see in the data, with no suggestion that this statistic is being
extended in its interpretation beyond the sample. While making
descriptive statements about the sample is always defensible, the true
goal is usually to try and learn something more fundamental to a larger
setting than merely the research participants at hand.

The true goal of discovering and/or describing fundamental truths
underlying data generation involves the use of \textbf{inferential
statistics}. Inferential statistics aim to extend observations from a
sample and generalize to the broader \textbf{population} from which the
sample was drawn. For example, if we are studying associations between
smoking and delay discounting, it is not particularly impactful to
merely summarize this association in 106 research participants. Instead,
we hope we have learned something that extends to the broader population
of smokers. A \textbf{Statistical Inference} is a statement made about
the broader population on the basis of data analyzed from a sample,
accompanied by an appropriate probability statement that quantifies
uncertainty.

\begin{figure}
\includegraphics[width=0.9\linewidth]{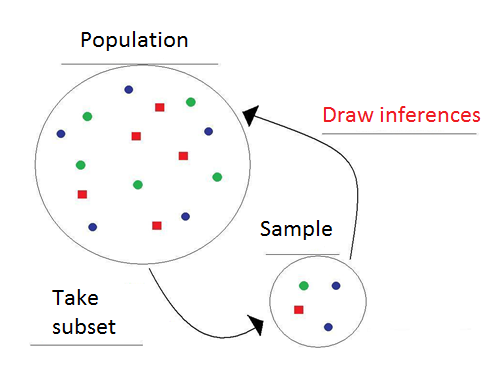} \caption{Statistical inference occurs when probability-based conclusions are drawn about the population on the basis of a carefully drawn and analyzed set of sample data.}\label{fig:figurename2}
\end{figure}

The above schema shows that there is an overall population of interest.
It is impractical or impossible to measure every unit in the population.
The major innovation of statistical reasoning is that, when a
\textbf{representative sample} is drawn from the population, statistical
methods (derived on the basis of probability theory) enable analysts to
generalize findings from the sample to the broader population with
accompanying statements of uncertainty. Uncertainty is explained using
the language of probability.

We will discuss three widely used inferential approaches here: point
estimation, confidence intervals, and hypothesis testing.

Point estimation occurs when the analyst purports that a summary
statistic is a ``good guess'' for the corresponding population value.
Using the above example, the sample mean for \(\text{ln}(k)\) among
smokers is -4.90. This is a summary statistic based on the sample.
However, the moment we say something like ``Our best guess for the true
population mean \(\text{ln}(k)\) is -4.90'' we engage in \emph{point
estimation} as a statistical inference exercise. The notion that a
sample statistic is ``close to'' or ``a good guess for'' the true but
unknown value in the population (of all smokers in this example) is
intuitively appealing and also mathematically justified. Upper level and
graduate courses in statistics use calculus and probability theory to
justify formally, for example, , for example that the sample mean is a
good guess for the population mean.

Confidence intervals are derived to ensure that true but unknown
population parameters are contained within the interval with a
user-controlled long run rate of success (e.g., we expect 95 out of 100
95\% confidence intervals to contain the true but unknown parameter).
Hypothesis tests are derived to have user controlled rates of obtaining
a false positive finding (i.e., type I error), traditionally set to 5\%
(i.e., \(\alpha=0.05\)). A good book that develops ideas of statistical
inference using calculus while teaching probability theory is (Wackerly,
Mendenhall, and Scheaffer 2014).

Of course, not just any sample is \textbf{representative} of the broader
population. The gold standard approach for obtaining a representative
sample is to obtain a \textbf{random sample}. The term ``random'' has a
specific technical meaning in this context. We do not mean
``haphazard,'' ``chaotic,'' or any other colloquial use of the term
``random.''

Instead, random sampling occurs when each element of the population has
a chance of being included in the sample, and a probability-based
mechanism is in charge of drawing the sample. You could think about
drawing names out of a really big hat. A \textbf{Simple random sample}
is the most conceptually straightforward design, as it gives every
subset of a fixed size the same chance of being selected, thus every
member of the population has the same chance of being included in the
sample.

It is easy to discuss drawing a sample completely randomly from a larger
population and admiring the simple elegance and amazing theoretical
properties of statistical approaches when this is the case. Reality is
messier than this. It is not easy to obtain a representative sample. It
might not be possible to even list every element in the population (such
a list is called the \textbf{sampling frame}), let alone locate all
individuals sampled from that list and ensure every individual selected
by the researcher actually enrolls in and completes the study.

I am quite fond of the following ``Fundamental Rule for using data for
inference,'' as it comments on the viability of applying statistical
methods to data even in situations where sampling is not perfect (Utts
and Heckard 2014). This rule is written as follows:

``The fundamental rule for using data for inference is that available
data can be used to make inferences about a much larger group if the
data can be considered to be representative with regard to the questions
of interest.''

Samples that are not drawn representatively from a population are
broadly known as \textbf{convenience samples}. Convenience samples can
be valuable (e.g., when studying an extremely rare disease, researchers
may consider themselves lucky to obtain data from anyone with the
disease regardless of sampling design). However, when convenience
samples are drawn and their data analyzed, we do have a certain level of
skepticism in the statistical results owing to the potential for
unquantifiable bias to manifest in the results.

Another key point is that representativeness is a function of the
sampling mechanism, not the sample size. An ideally drawn random sample
is representative of the population even if the sample size is not
large. By contrast, if a flawed sampling design specifically excludes
certain segments of the population (e.g., affluent recreational cocaine
users are unlikely to enroll in a discounting study on cocaine users)
then merely increasing the sample size will not overcome bias introduced
by the flawed sampling approach. Sampling theory is its own field within
statistics. A good introductory textbook on sampling is (Lohr 2021).

The notion of randomization as a method to ensure valid inference is
also ubiquitous within the study of the design of experiments. Unlike in
an observational study, in an \textbf{experiment}, the researcher
assigns experimental units to different treatments. Sophisticated
experimental designs can randomize in a way to account for known
confounds (e.g., ensure that members of each socioeconomic stratum are
present in all experimental groups so socioeconomic effects do not mask
experimental effects). Even if many known confounding variables are
accounted for in this way, randomization remains essential. \emph{Random
assignment of units to treatments is ingenious because randomization
makes treatment groups the same on average, both with respect to known
and also with respect to unknown potential confounding variables.} A
good book for further reasoning design of experiments is (Montgomery
2008).

\textbf{Executing Stage 2 analysis using inferential statistics }

The exploratory data analysis above is descriptive in nature. We have
not (yet) made any assertions that the patterns in these data extend to
a broader population. Next, we will do just this for the three
comparisons of discounting rates (i) between males and females, (ii)
between smokers and non-smokers, and (iii) as a function of age. We will
assume our sample data are representative of a relevant larger
population. We will illustrate hypothesis tests and confidence intervals
for parameters for each comparison. Like many topics in this chapter, we
provide a brief overview of common techniques appropriate for analysis
in this setting. We assume the reader has been exposed to the basic idea
of statistical hypothesis testing and confidence intervals. For those
who may wish to review these concepts, an excellent book to further
study the applied statistical techniques presented here is (Utts and
Heckard 2014).

\hypertarget{comparing-discounting-between-males-and-females}{%
\section{Comparing discounting between males and
females}\label{comparing-discounting-between-males-and-females}}

Recall the box plot of estimated \(\text{ln}(\hat{k})\) values as a
function of sex.

\begin{Shaded}
\begin{Highlighting}[]
\FunctionTok{boxplot}\NormalTok{(lnk}\SpecialCharTok{\textasciitilde{}}\NormalTok{dat}\SpecialCharTok{$}\NormalTok{gender,}\AttributeTok{xlab=}\StringTok{"Gender"}\NormalTok{)}
\end{Highlighting}
\end{Shaded}

\includegraphics{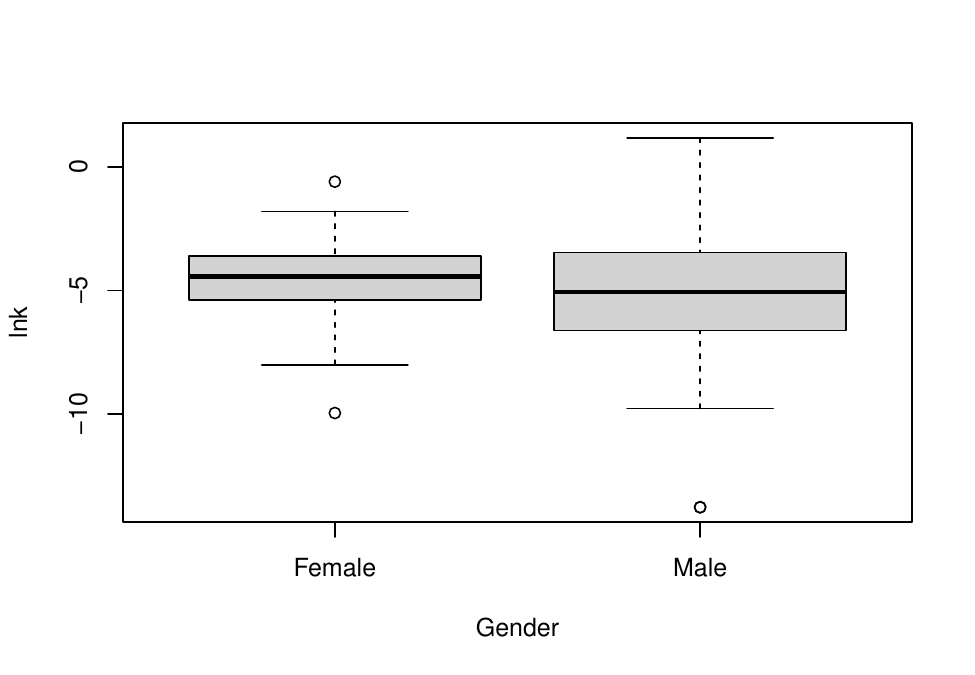}

\begin{Shaded}
\begin{Highlighting}[]
\FunctionTok{t.test}\NormalTok{(lnk}\SpecialCharTok{\textasciitilde{}}\NormalTok{dat}\SpecialCharTok{$}\NormalTok{gender)}
\end{Highlighting}
\end{Shaded}

\begin{verbatim}
## 
##  Welch Two Sample t-test
## 
## data:  lnk by dat$gender
## t = 0.66633, df = 64.897, p-value = 0.5076
## alternative hypothesis: true difference in means between group Female and group Male is not equal to 0
## 95 percent confidence interval:
##  -0.6400485  1.2809533
## sample estimates:
## mean in group Female   mean in group Male 
##            -4.665273            -4.985726
\end{verbatim}

The above code compares males and females in terms of discounting rate.
The box plot shows little shift between the genders, although it
visually appears that that males are generating more variability in
their discounting rates. We estimate the average discounting
\(\text{ln}(\hat{k})\) among females and males to be -4.67 and -4.99,
respectively. The t.test() code conducts a statistical comparison
between these groups. The ``Welch Two Sample t-test'' phrase indicates
that the testing procedure does not assume common variance between the
groups. For the test with null hypothesis of no difference in
\(\text{ln}(k)\) between the groups, the p-value is 0.508. We fail to
reject the null hypothesis, as results as or more extreme against the
null would arise half the time if the null hypothesis is true. (I.e., If
there truly is no underlying difference between groups, results such as
those we see here are not particularly improbable.) A 95\% confidence
interval for the difference in means is (-0.64, 1.28). Since this
interval contains zero, it is plausible that there is no underlying
difference between males and females on the basis of these data.

Using the data in this chapter as an example, we might study the
association between smoking status and delay discounting as measured by
\(\text{ln}(\hat{k})\) within this sample, which contains 52 non-smokers
and 54 smokers. We can accomplish this graphically with box plots and in
tabular format with summary statistics.

\hypertarget{comparing-discounting-between-smokers-and-non-smokers}{%
\section{Comparing discounting between smokers and
non-smokers}\label{comparing-discounting-between-smokers-and-non-smokers}}

Recall the box plot of estimated \(\text{ln}(\hat{k})\) values as a
function of smoking status.

\begin{Shaded}
\begin{Highlighting}[]
\FunctionTok{boxplot}\NormalTok{(lnk}\SpecialCharTok{\textasciitilde{}}\NormalTok{dat}\SpecialCharTok{$}\NormalTok{smoke\_cigs,}\AttributeTok{xlab=}\StringTok{"Smoking status"}\NormalTok{)}
\end{Highlighting}
\end{Shaded}

\includegraphics{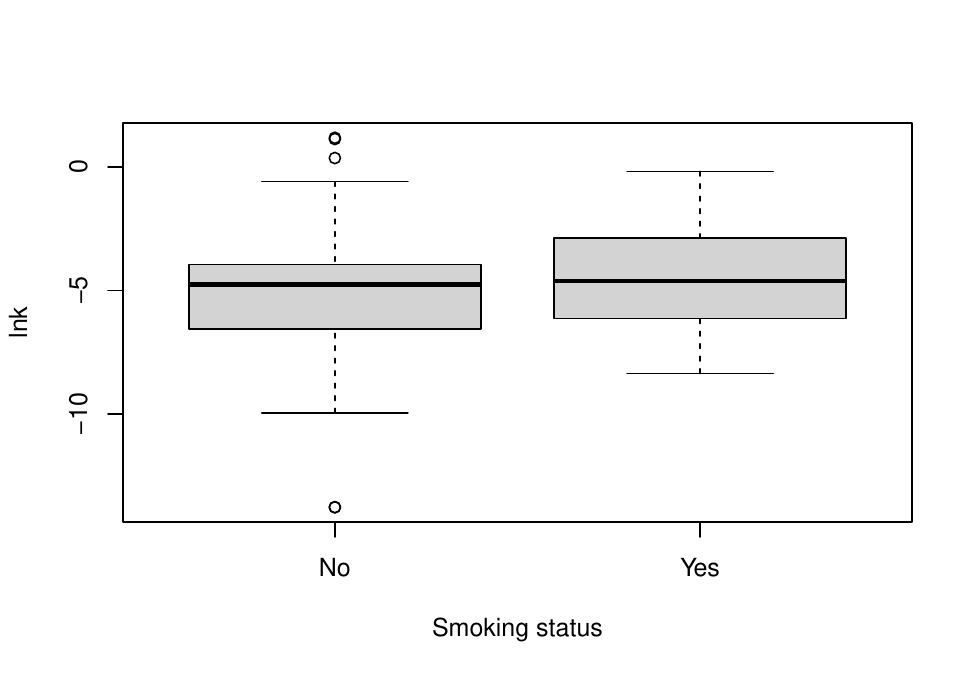}

\begin{Shaded}
\begin{Highlighting}[]
\FunctionTok{t.test}\NormalTok{(lnk}\SpecialCharTok{\textasciitilde{}}\NormalTok{dat}\SpecialCharTok{$}\NormalTok{smoke\_cigs)}
\end{Highlighting}
\end{Shaded}

\begin{verbatim}
## 
##  Welch Two Sample t-test
## 
## data:  lnk by dat$smoke_cigs
## t = -0.96964, df = 95.021, p-value = 0.3347
## alternative hypothesis: true difference in means between group No and group Yes is not equal to 0
## 95 percent confidence interval:
##  -1.4765591  0.5074967
## sample estimates:
##  mean in group No mean in group Yes 
##         -5.150938         -4.666407
\end{verbatim}

The box plot above suggests that smokers may have higher discounting
rates on average than non-smokers. We estimate the average discounting
\(\text{ln}(\hat{k})\) among nonsmokers is -5.15 and among smokers the
average discounting \(\text{ln}(\hat{k})\) is estimated to be -4.67
among non-smokers. For the test with null hypothesis of no difference in
\(\text{ln}(k)\) between the groups, the p-value is 0.334. We fail to
reject the null hypothesis, as results as or more extreme against the
null would arise one third of the time if the null hypothesis is true. A
95\% confidence interval for the difference in means is (-1.48, 0.51).
Since this interval contains zero, it is plausible that there is no
underlying difference in discount rate between smokers and non-smokers
on the basis of these data.

\emph{Assessing discounting as a function of age}

Recall the scatter plot with \(\text{ln}(\hat{k})\) on the vertical axis
and age on the horizontal axis. The following code fits a \emph{simple
linear regression model} to these data. Simple linear regression
proceeds by identifying the straight line that is ``closest'' to the
data in a least squares sense. We have already seen the idea of least
squares with the nonlinear regression approach we used to fit the Mazur
function to indifference point data. As before, the least squares
approach identifies regression lines for which the sum of squared
residuals is minimized. In this simple linear model, the parameters are
the slope and y-intercept of the regression line.

\begin{Shaded}
\begin{Highlighting}[]
\FunctionTok{plot}\NormalTok{(dat}\SpecialCharTok{$}\NormalTok{age,lnk)}
\NormalTok{mod}\OtherTok{\textless{}{-}}\FunctionTok{lm}\NormalTok{(lnk}\SpecialCharTok{\textasciitilde{}}\NormalTok{dat}\SpecialCharTok{$}\NormalTok{age)}
\FunctionTok{abline}\NormalTok{(mod,}\AttributeTok{col=}\StringTok{\textquotesingle{}red\textquotesingle{}}\NormalTok{,}\AttributeTok{xlab=}\StringTok{"age"}\NormalTok{,}\AttributeTok{lwd=}\DecValTok{2}\NormalTok{)}
\end{Highlighting}
\end{Shaded}

\includegraphics{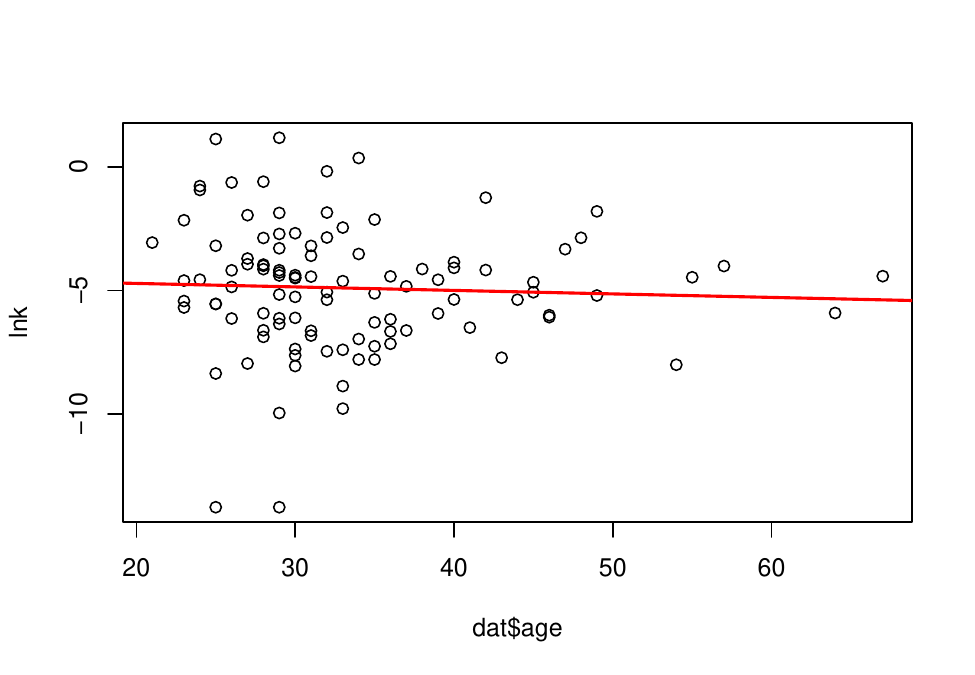}

\begin{Shaded}
\begin{Highlighting}[]
\FunctionTok{summary}\NormalTok{(mod)}
\end{Highlighting}
\end{Shaded}

\begin{verbatim}
## 
## Call:
## lm(formula = lnk ~ dat$age)
## 
## Residuals:
##     Min      1Q  Median      3Q     Max 
## -8.9914 -1.4603  0.1821  1.3617  6.0244 
## 
## Coefficients:
##             Estimate Std. Error t value Pr(>|t|)    
## (Intercept) -4.42937    0.99637  -4.446  2.2e-05 ***
## dat$age     -0.01418    0.02880  -0.492    0.624    
## ---
## Signif. codes:  0 '***' 0.001 '**' 0.01 '*' 0.05 '.' 0.1 ' ' 1
## 
## Residual standard error: 2.567 on 104 degrees of freedom
## Multiple R-squared:  0.002323,   Adjusted R-squared:  -0.00727 
## F-statistic: 0.2422 on 1 and 104 DF,  p-value: 0.6237
\end{verbatim}

\begin{Shaded}
\begin{Highlighting}[]
\FunctionTok{confint}\NormalTok{(mod)}
\end{Highlighting}
\end{Shaded}

\begin{verbatim}
##                   2.5 %      97.5 %
## (Intercept) -6.40520508 -2.45353045
## dat$age     -0.07129443  0.04294416
\end{verbatim}

The estimated slope and intercept are -0.014 and -4.429, respectively.
This means that for a one year increase in age, we estimate the average
\(\text{ln}(k)\) \emph{decreases} (due to negative sign) by 0.014. The
mathematical interpretation of the y-intercept is that individuals with
age=0 have an average \(\text{ln}(k)\) of -4.43. In this case, it is not
meaningful to consider a zero-year-old's discounting, so direct
interpretation of the y-intercept is basically meaningless. Nonetheless,
the y-intercept is important because it governs the overall height of
the line. Don't omit the y-intercept from the model unless you have
compelling scientific information that at x=0 the outcome y=0 (e.g., x
is fuel burned and y is exhaust emitted.)

As before, we may be interested in confidence intervals and hypothesis
tests for the parameters of interest. For reasons we just described, we
focus on interpreting these for the slope parameter in this analysis.
The summary() function provides estimates, standard errors, test
statistics, and p-values for the null hypothesis that the corresponding
parameter value is zero. In this case, for the age effect, the p-value
is 0.624. We fail to reject the null hypothesis because a slope value of
zero is plausible on the basis of these observed data. This corresponds
to the visual evidence in the scatter plot where these data do not
appear to show much of a linear relationship. A 95\% confidence interval
for the slope is (-0.071, 0.043).

Ordinary regression assumes that residuals have a normal distribution
with constant variance. Some of the exercises at the end of this chapter
will guide you through a way to assess the plausibility of these
assumptions.

Regression is one of the most flexible and important classes of
statistical techniques. Regression can be tailored to suit a vast
collection of problems. Further topics include multiple regression
(i.e., what to do when multiple predictors are available), model
selection (i.e., which predictors should be included and what function
of these best describe the response), regression for outcomes that do
not have normally distributed residuals (i.e., \emph{generalized linear
models} are useful for outcomes with categorical or count outcomes). A
good book to learn more about regression is (Montgomery, Peck, and
Vining 2015).

\textbf{Contemporary issues in statistical practice}

There are a number of interesting issues and controversies related to
contemporary statistical practice. The largest-looming of these is the
\emph{replication crisis}. If the goal of science is to uncover and
understand observable truths in the universe, then similar studies
should produce similar results. If a well-executed replication study
fails to produce the same conclusions as the original study, then
arguably the original study fell short of finding the correct answer to
the research question. Readers in this area are hopefully aware of the
replication crisis in general (Ioannidis 2005) and in psychology
specifically (Collaboration 2015). Many factors can contribute to the
failure of a study to replicate, including choices made during
experimental design, data collection and processing, initial exploratory
data analysis and modeling. Certainly, a lot of scrutiny has been given
to issues of classical hypothesis testing, including the use of p-values
and surrounding practices (Leek and Peng 2015). (S. S. Young and Karr
2011) describes the situation of scientific publishing based on
statistical significance and p-values as an out-of-control process. In
response to these concerns, the American Statistical Association issued
a statement with clarifications and a few suggestions to clarify
definitions and best practices related to p-values (Wasserstein and
Lazar 2016). The literature review in (Franck, Madigan, and Lazar 2022)
describes some of the conversations about statistical hypothesis testing
and p-values that have been tied to the replication crisis. An entire
chapter or possibly even a textbook could be filled with this
discussion. The goal here is to outline some issues broadly and describe
statistical workflows that are transparent and responsible.

\textbf{The previous textbook chapter analyses are not a blueprint for a
reasonable analytic strategy in a research project.} Our goal has been
to demonstrate a wide variety of relevant methods for analyzing
discounting data. With this focus, we have not articulated a primary
analysis strategy. We have not guarded against problems of multiple
comparison, i.e., \emph{multiplicity}. We have not decided in advance
how to handle participants who fail attention checks or commit Johnson
\& Bickel violations. Instead, we have conducted a collection of
individual analyses to illustrate each concept in a safe sandbox using
data without particularly strong associations between discounting and
other measures. We make no claims about research conclusions on the
basis of the analyses in this chapter. This has been a training ground,
and the next exercise is to think more about how a reasonable analytic
pipeline will look for a research project.

Behavior analysis know well that organisms tend to behave in a manner
for which they are rewarded. This extends to scientific publishing as
well, where the rewards to publish novel, ``statistically significant''
research are compelling. These include rising notoriety, competitive
advantage in the pursuit of extramural funding, and the chance to
continue participating in science. The consequences for failing to
publish novel research are punishing and involve either a quick or
prolonged exit from a scientific career.

Mix this reward/punishment structure with the historical but
controversial bright line rule of declaring any scientific finding to be
``statistically significant'' so long as the data analysis yields a
p-value below 0.05, and there is an inherent conflict of interest that
incentivizes analysts to select analyses that achieve ``statistical
significance'' and portraying these results in a manner that failed to
acknowledge uncertainty in conclusions. This process of repeatedly
conducting analyses and/or gathering more data, then stopping once an
analysis has a p-value below \(\alpha=0.05\) is known as p-hacking, and
it is problematic because it fundamentally violates the statistical
error rates that come with hypothesis testing.

By way of review, classical hypothesis testing works by controlling the
rate at which a false positive finding occurs. This false positive error
rate, also known as the Type I error rate, is set in advance and has
traditionally been \(\alpha=0.05\). This means that if even in the case
where every other aspect of the study is done correctly, we would reject
the null hypothesis in 5\% of the cases where it is actually true, i.e.,
make false positive error. Note that this traditional threshold has
recently been challenged and there are reasonable arguments for setting
\(\alpha=0.005\) for new discoveries (Benjamin et al. 2018). Thus, the
Type I error rate \(\alpha\) is the rate at which an analyst is willing
to incorrectly reject a true null hypothesis (which is commonly a
hypothesis of no association although specific null values of parameters
can be stipulated).

The idea is that hypothesis testing protects the broader scientific
effort from a preponderance of false positive findings. But if an
unscrupulous analyst decides to continually try various analysis plans
until a single analysis yields a p-value below 0.05, then obviously this
procedure leads to much higher than a 5\% false positive rate.

This conflict of interest can be insidious. Imagine your primary
analysis plan calls for using Mazur's \(\text{ln}(k)\) as a discounting
metric and examining its association with a key predictor. The
statistical analysis reveals a small association between your predictor
and discounting, with a p-value of 0.13. According to this primary
analysis (and a traditional error rate of \(\alpha=0.05\)), the true
association is plausibly null, your chances of publishing in a
high-impact journal are low, you have no compelling pilot data for a
grant application, and the clock is ticking on the remaining time in
your graduate school/post-doc/faculty position.

Imagine you decide to mess around a little more to get a lower p-value.
Suppose switching to a different metric (e.g., AUC) and eliminating two
outliers makes the same analysis yield a p-value of 0.03. Now, the
actual false positive error rate of your procedure is MUCH higher than
5\%. You must choose between fundamentally violating the error rate and
publishing your work in a less-than-honest manner (presenting the
results as ironclad), or possibly risking your future in academia if you
publish nothing.

The term ``p-hacking'' refers to deliberately choosing specific analyses
that reveal ``statistically significant'' associations while
deliberately failing to acknowledge that a greater number of statistical
tests were conducted whose results were non-significant, and the choice
of which analysis to report was based only on finding an analysis with a
significant p-value. The incentive to p-hack is that by illegitimately
obtaining low p-values, research findings are presented as though they
are more ``statistically significant'' than the data suggest, more
journals are more likely to look favorably on the findings, and the
researcher is able to reap the rewards of successful publication, albeit
at the cost of polluting the literature with findings that are unlikely
to replicate, potentially sending other researchers down faulty lines of
inquiry.

One idea to combat this incentive structure is to pursue the idea of
preregistered journal articles. The basic idea with preregistered
articles is that a researcher will draft the introduction and methods
section (which fully details the primary analysis plan) for a paper,
then submit these to a journal for publication \emph{before} collecting
and analyzing the data. The journal accepts or rejects the paper based
on scientific novelty and methodological soundness. Then the researcher
completes the study and and is guaranteed publication regardless of the
results of the statistical analysis. If an investigation is deemed
scientifically sound but does not yield strong evidence of association
for a reasonably powered study, then that should be reported.

Here are three pieces of advice we'd offer to aspiring statistical
analysts in the world of behavior analysis:

\begin{enumerate}
\def\labelenumi{(\arabic{enumi})}
\tightlist
\item
  Pre-plan your primary analysis. Describe everything from data
  collection, articulation of inclusion/exclusion criteria, how to
  handle outliers, how missing data will be handled, which models will
  be fit, how associations between discounting will be quantified, and
  which multiplicity adjustments will be used BEFORE data collection
  begins. Execute that plan and report the results.
\item
  Transparently report all comparisons and secondary analyses you
  conduct. Of course, data may provide value and insights beyond what
  the analyst anticipated with their primary analysis plan. (Imagine if
  Alexander Fleming threw away the contaminated Petri dish that led to
  the discovery of antibiotics!) By clearly demarcating primary analyses
  and secondary analyses, and reporting on all analyses attempted, you
  acknowledge multiplicity and protect against perceptions of p-hacking.
  (Relegate low-impact analysis results to an appendix or online
  supplement.)
\item
  Report and discuss interpretable effect sizes and graphical displays
  of data. Help the audience understand the \emph{strength of the
  associations} among key variables in your data. Do not merely report
  p-values and declare ``statistical significance.'' A good reference is
  (Jacob 2007).
\end{enumerate}

Let us also briefly distinguish between replication and reproducibility.
A study is \emph{replicable} if another research group can redo your
study by following your study protocol and obtain the same results and
conclusions as your study. \emph{Reproducibility} refers to the ability
of an analyst with access to your data and code to reproduce the same
results you published on the basis of your data. It is important for
analysts to be impeccably organized in data collection and store all
analysis codes for posterity. Increasingly, journals urge authors to
make raw data and analysis scripts public where possible so that claims
in articles can be reproduced by other researchers. More than once in my
career I have received requests from researchers for such materials.

Finally, we note that the list of things ``not-to-do'' is infinity
large. Once upon a time at Virginia Tech, the consulting director asked
the design of experiments teacher why nobody has written a book of
common mistakes in study design. To this, the professor responded that
such a book could never be completed!

\hypertarget{the-role-of-planning-in-scientific-investigation}{%
\subsection{The role of planning in scientific
investigation}\label{the-role-of-planning-in-scientific-investigation}}

We have meandered through a number of different topics related to the
analysis of delay discounting data. In order for this chapter to be
maximally informative, we have introduced a topic, illustrated it on our
example data set, and then explained the outcome of that investigation.
We have had little regard for what topic is next, how many (i.e., the
\textbf{multiplicity} of) statistical comparisons were conducted, or
establishing a concrete primary analysis plan. We believe this is the
way to maximize the educational value of the chapter.

In many ways this is the \emph{opposite} of what should be done when
conducting a formal research study. Before one begins analyzing data, we
remind the reader that \emph{this has been an instructional textbook
chapter not intended to resemble a typical research analysis.} This
distinction is important, because when designing a research study, the
analyst must consider many study design issues \emph{before data
analysis begins}. Some of the specific statistical and data analytic
issues are:

\begin{enumerate}
\def\labelenumi{(\roman{enumi})}
\tightlist
\item
  Missing data. The reason for the patterns of missing data guides best
  practices for subsequent analysis. (Little and Rubin 2014; Allison
  2001).
\item
  Whether and how to pre-screen participant data and how to handle data
  patterns that diverge sharply from model form. (e.g., we expect a
  decreasing trend between valuation and delay but sometimes see a flat,
  wildly variable, or increasing trend which seems irrational).
\item
  How to handle outliers and other unusual data points. Even a small
  number of extreme data points can greatly influence analysis results.
  Why are there unusual points and how should they be handled?
\item
  Which models to fit and which methods to fit them. Many competing
  models have been proposed.
\item
  How many comparisons will we make and how to make them? (E.g. smokers
  versus non smokers? Number of cigarettes smoked daily?). Failure to
  account for the effects of multiple testing (i.e., multiplicity) leads
  to greatly increased chances of false positive findings. Deliberately
  exploiting this practice is known as ``p-hacking.'' More discussion of
  some of these issues and other contemporary issues and controversies
  related to statistical analysis can be found in the
  \textbf{Contemporary issues in statistical practice} Section above.
\end{enumerate}

To summarize the point, the textbook chapter is designed to impart
skills and knowledge one item at a time. By contrast, research studies
are proactively planned by research teams with combined experience in
statistical, behavioral analysis, and other relevant research skills and
knowledge.

\textbf{Continuing your statistical training}

This chapter is intended to be an on-ramp for behavior analysts who wish
to move towards greater statistical knowledge. So unfortunately, if we
were successful you may feel you are suddenly on a highway with little
idea of what to do next! The machinery here is powerful enough to be
dangerous, and we share the statistical highway with hobbyists,
professionals, fools, and charlatans. What we have covered here is just
a tiny introductory piece of the broader statistical field.

To continue the driving example, the best rule is to move carefully and
safely at the speed you are comfortable. If you aren't sure what to do
next, read! Ask! Like any road trip, the more knowledge you have about
the route ahead, the better. Consider the value in finding trusted
traveling companions who can help navigate and drive through challenging
parts of the landscape. The biggest surprise about statistics, to me, is
how subtle the field is. I think this is mostly due to the fact humans
are not naturally good at understanding uncertainty, and statistics is
all about finding a few crude signals is a world dominated by
uncertainty. We wish the reader the best of luck on this journey, and
don't hesitate to reach out.

\textbf{Exercises}

General questions:

\begin{enumerate}
\def\labelenumi{(\arabic{enumi})}
\tightlist
\item
  Let's review the logarithm function more thoroughly. Do you know what
  the logarithm function looks like? Take a pen and paper and try to
  draw the function \(y=\text{ln}(x)\). Now, write an R program to graph
  this function. In what ways is your drawing similar to the R plot? Are
  there any differences or things you learned from this activity?
\end{enumerate}

Stage 1 questions:

\begin{enumerate}
\def\labelenumi{(\arabic{enumi})}
\setcounter{enumi}{1}
\tightlist
\item
  Consider again the first stage of analysis. We used the Mazur model
  but another discounting model that has two parameters was popularized
  by Howard Rachlin (Rachlin 2006):
\end{enumerate}

\[E(y)=\frac{A}{1 + k*D^s},\]

where \(k\) and \(s\) are unknown parameters that must be estimated.
Adjust the code that fits the Mazur model to instead fit the Rachlin
model to the first participant's data. Report the estimated \(\hat{k}\)
and \(\hat{s}\) values from the Rachlin model, and plot indifference
points by delay including regression lines for both Mauzr's and
Rachlin's model.

\begin{enumerate}
\def\labelenumi{(\arabic{enumi})}
\setcounter{enumi}{2}
\item
  Recall that Effective Delay 50 (ED50) is the delay at which a future
  reward is devalued by half of the larger later amount. Use algebra to
  show that the ED50 for the Rachlin model is
  \[\Big(\frac{1}{k}\Big)^{1/s}.\] \emph{Hint - set right hand size of
  Rachlin equation in the first problem to \(A/2\), then solve for
  \(D\).}
\item
  Estimate ED50 for the first participant's data using both the Mazur
  model and also the Rachlin model. Do these estimates agree to a
  reasonable extent? Does this corroborate what you see in the plot from
  the previous problem? Do you think the agreement between models would
  be as similar for ED90 (i.e., the delay at which 90\% of the larger
  later reward is devalued)? Justify your explanation using the plot
  from problem (2).
\end{enumerate}

Stage 2 questions

\begin{enumerate}
\def\labelenumi{(\arabic{enumi})}
\setcounter{enumi}{4}
\item
  Report the number of participants who violated one or both Johnson \&
  Bickel criteria (i.e., they have a value of JBviol=1 in the original
  data). Report the number of participants who failed the attention
  check (i.e., they have a value of ddattend=``\$0.00 now'' in the
  original data). Make a two-way table of these variables and comment on
  the extent to which those who failed the attention check are also
  Johnson \& Bickel criteria violators.
\item
  Consider an analysis in which violators of the Johnson \& Bickel
  criteria are not analyzed with the full group in the broader
  statistical analysis. Subset the data to include only individuals who
  do not violate the criteria (i.e., they have a value of JBviol=0 in
  the original data). Redo the two-stage analyses presented in the
  chapter to determine whether our results comparing discounting between
  smokers and non-smokers, females and males change, and whether the
  association between discounting and age change if Johnson \& Bickel
  criteria violators are omitted from the main analysis.
\item
  Write a loop that computes and stores AUC for all subject
  participants. Make a histogram of these AUC values. Take the natural
  log of each AUC value and make a second histogram of these lnAUC
  values. Which of AUC and logAUC do you think more closely resembles a
  normal distribution and why?
\item
  Create a pairs plot that includes Mazur lnk, Mazur ED50, AUC, nnd
  ln(AUC). Interpret the plot. Does this plot alone definitively
  determine which discounting metric is best?
\item
  Ordinary regression assumes residuals (i.e.~the difference between
  data points and the corresponding values on the regression line at the
  same x values) are normally distributed, and that variance in these
  residuals is constant. For the regression problem with
  \(\text{ln}(\hat{k})\) as the outcome and age as the predictor, use
  the skills in this book plus a few Google searches to (i) Obtain the
  predicted values and residuals from this analysis, (ii) make a
  histogram of the residuals, and (iii) a scatter plot with predicted
  values on the horizontal axis and residuals on the vertical axis. Does
  it appear that the assumptions are reasonably satisfied?
\end{enumerate}

\textbf{Hints for Exercises}

Unless otherwise stated, objects in the code below are as defined in the
chapter.

\hypertarget{section}{%
\subsection{(1)}\label{section}}

\begin{Shaded}
\begin{Highlighting}[]
\CommentTok{\#Log is defined on positive number line}
\CommentTok{\#Make a dense grid of points there}
\NormalTok{x}\OtherTok{\textless{}{-}}\FunctionTok{seq}\NormalTok{(.}\DecValTok{001}\NormalTok{,}\DecValTok{4}\NormalTok{,.}\DecValTok{001}\NormalTok{)}
\NormalTok{lnx}\OtherTok{\textless{}{-}}\FunctionTok{log}\NormalTok{(x)}
\FunctionTok{plot}\NormalTok{(x,lnx,}\AttributeTok{type=}\StringTok{\textquotesingle{}l\textquotesingle{}}\NormalTok{)}
\FunctionTok{abline}\NormalTok{(}\AttributeTok{v=}\DecValTok{1}\NormalTok{,}\AttributeTok{col=}\StringTok{\textquotesingle{}lightgrey\textquotesingle{}}\NormalTok{,}\AttributeTok{lty=}\DecValTok{3}\NormalTok{)}
\FunctionTok{abline}\NormalTok{(}\AttributeTok{h=}\DecValTok{0}\NormalTok{,}\AttributeTok{col=}\StringTok{\textquotesingle{}lightgrey\textquotesingle{}}\NormalTok{,}\AttributeTok{lty=}\DecValTok{3}\NormalTok{)}
\end{Highlighting}
\end{Shaded}

\includegraphics{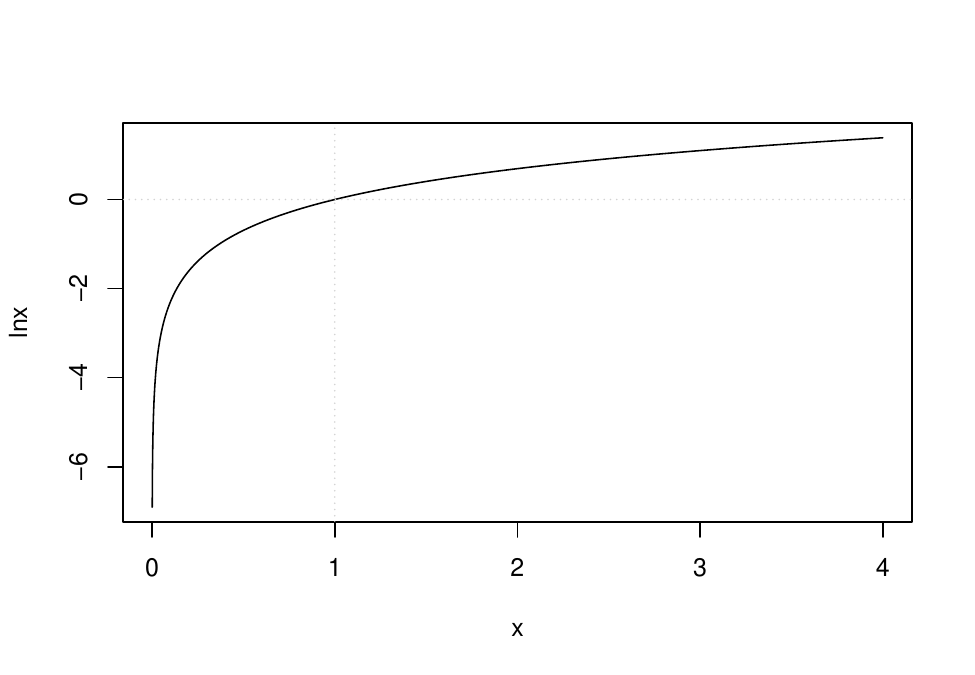}

\begin{Shaded}
\begin{Highlighting}[]
\CommentTok{\#log of 1 is zero}
\CommentTok{\#log of zero not defined }
\CommentTok{\#vertical range of log function as x increases is unbounded}
\CommentTok{\#smooth, monotone increasing function}
\CommentTok{\#not defined for negative numbers}
\end{Highlighting}
\end{Shaded}

\hypertarget{section-1}{%
\subsection{(2)}\label{section-1}}

\begin{Shaded}
\begin{Highlighting}[]
\NormalTok{i}\OtherTok{\textless{}{-}}\DecValTok{1} \CommentTok{\#Set an index number to 1 for the first subject}
\NormalTok{y.frame}\OtherTok{\textless{}{-}}\NormalTok{dat[i,}\DecValTok{5}\SpecialCharTok{:}\DecValTok{11}\NormalTok{,drop}\OtherTok{=}\ConstantTok{FALSE}\NormalTok{] }\CommentTok{\#ith row, columns 5 through 11}
\NormalTok{y}\OtherTok{\textless{}{-}}\FunctionTok{as.vector}\NormalTok{(}\FunctionTok{as.matrix}\NormalTok{(}\FunctionTok{t}\NormalTok{(y.frame))) }\CommentTok{\#Make y a vector}
\NormalTok{D}\OtherTok{\textless{}{-}}\FunctionTok{c}\NormalTok{(}\DecValTok{1}\NormalTok{,}\DecValTok{7}\NormalTok{,}\DecValTok{30}\NormalTok{,}\DecValTok{90}\NormalTok{,}\DecValTok{365}\NormalTok{,}\DecValTok{1825}\NormalTok{,}\DecValTok{9125}\NormalTok{) }\CommentTok{\#Define the delays D}

\CommentTok{\#Fit the Rachlin model}
\NormalTok{Rachlin.mod}\OtherTok{\textless{}{-}}\FunctionTok{nls}\NormalTok{(y}\SpecialCharTok{\textasciitilde{}}\DecValTok{1}\SpecialCharTok{/}\NormalTok{(}\DecValTok{1}\SpecialCharTok{+}\NormalTok{k}\SpecialCharTok{*}\NormalTok{D}\SpecialCharTok{\^{}}\NormalTok{s),}\AttributeTok{start=}\FunctionTok{list}\NormalTok{(}\AttributeTok{k=}\NormalTok{.}\DecValTok{1}\NormalTok{,}\AttributeTok{s=}\NormalTok{.}\DecValTok{1}\NormalTok{)) }\CommentTok{\#Fit the Mazur model to these data}
\FunctionTok{summary}\NormalTok{(mod) }\CommentTok{\#Summary of the fitted model}
\end{Highlighting}
\end{Shaded}

\begin{verbatim}
## 
## Call:
## lm(formula = lnk ~ dat$age)
## 
## Residuals:
##     Min      1Q  Median      3Q     Max 
## -8.9914 -1.4603  0.1821  1.3617  6.0244 
## 
## Coefficients:
##             Estimate Std. Error t value Pr(>|t|)    
## (Intercept) -4.42937    0.99637  -4.446  2.2e-05 ***
## dat$age     -0.01418    0.02880  -0.492    0.624    
## ---
## Signif. codes:  0 '***' 0.001 '**' 0.01 '*' 0.05 '.' 0.1 ' ' 1
## 
## Residual standard error: 2.567 on 104 degrees of freedom
## Multiple R-squared:  0.002323,   Adjusted R-squared:  -0.00727 
## F-statistic: 0.2422 on 1 and 104 DF,  p-value: 0.6237
\end{verbatim}

\begin{Shaded}
\begin{Highlighting}[]
\NormalTok{k.hat}\OtherTok{\textless{}{-}}\FunctionTok{summary}\NormalTok{(Rachlin.mod)}\SpecialCharTok{$}\NormalTok{coef[}\DecValTok{1}\NormalTok{,}\DecValTok{1}\NormalTok{]}
\NormalTok{s.hat}\OtherTok{\textless{}{-}}\FunctionTok{summary}\NormalTok{(Rachlin.mod)}\SpecialCharTok{$}\NormalTok{coef[}\DecValTok{2}\NormalTok{,}\DecValTok{1}\NormalTok{]}
\FunctionTok{print}\NormalTok{(}\StringTok{"k.hat and s.hat"}\NormalTok{)}
\end{Highlighting}
\end{Shaded}

\begin{verbatim}
## [1] "k.hat and s.hat"
\end{verbatim}

\begin{Shaded}
\begin{Highlighting}[]
\NormalTok{k.hat; s.hat}
\end{Highlighting}
\end{Shaded}

\begin{verbatim}
## [1] 9.418193e-05
\end{verbatim}

\begin{verbatim}
## [1] 1.277788
\end{verbatim}

\begin{Shaded}
\begin{Highlighting}[]
\NormalTok{D.s}\OtherTok{\textless{}{-}}\FunctionTok{seq}\NormalTok{(}\DecValTok{0}\NormalTok{,}\DecValTok{9500}\NormalTok{,}\DecValTok{1}\NormalTok{) }\CommentTok{\#Create a fine grid across delays. Used later to plot the regression line.}
\NormalTok{preds.Rachlin}\OtherTok{\textless{}{-}}\FunctionTok{predict}\NormalTok{(Rachlin.mod,}\AttributeTok{newdata=}\FunctionTok{data.frame}\NormalTok{(}\AttributeTok{D=}\NormalTok{D.s)) }\CommentTok{\#Obtain predicted values }
\FunctionTok{plot}\NormalTok{(D,y,}\AttributeTok{xlab=}\StringTok{\textquotesingle{}Delay (days)\textquotesingle{}}\NormalTok{,}\AttributeTok{ylab=}\StringTok{\textquotesingle{}Indifference point\textquotesingle{}}\NormalTok{,}
     \AttributeTok{main=}\StringTok{"Indifference points, model fit, and ED50"}\NormalTok{) }\CommentTok{\#Scatter plot}
\FunctionTok{lines}\NormalTok{(D.s,preds.Rachlin,}\AttributeTok{col=}\StringTok{\textquotesingle{}blue\textquotesingle{}}\NormalTok{) }\CommentTok{\#Add the regression line to the plot}
\FunctionTok{lines}\NormalTok{(D.s,preds,}\AttributeTok{col=}\StringTok{\textquotesingle{}red\textquotesingle{}}\NormalTok{,}\AttributeTok{lty=}\DecValTok{2}\NormalTok{) }\CommentTok{\#Add the regression line to the plot}
\FunctionTok{legend}\NormalTok{(}\DecValTok{6000}\NormalTok{,}\FloatTok{0.8}\NormalTok{,}\AttributeTok{legend=}\FunctionTok{c}\NormalTok{(}\StringTok{"Mazur"}\NormalTok{,}\StringTok{"Rachlin"}\NormalTok{),}\AttributeTok{col=}\FunctionTok{c}\NormalTok{(}\StringTok{"red"}\NormalTok{,}\StringTok{"blue"}\NormalTok{),}\AttributeTok{lty=}\FunctionTok{c}\NormalTok{(}\DecValTok{2}\NormalTok{,}\DecValTok{1}\NormalTok{))}
\end{Highlighting}
\end{Shaded}

\includegraphics{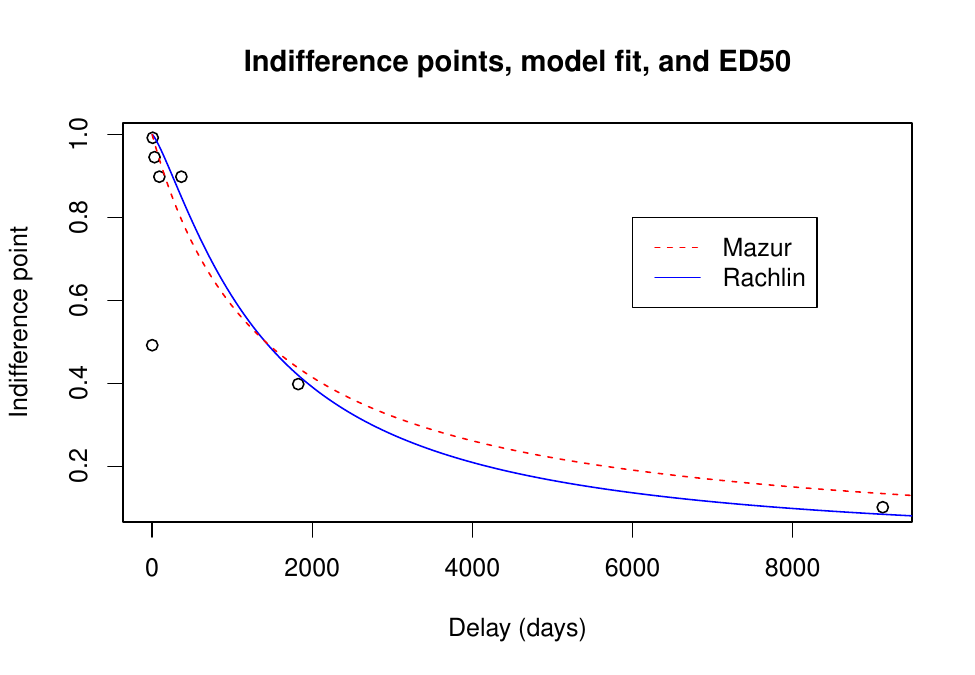}

The parameter estimates are \(\hat{k}= 9.4 \times 10^{-5} = .000094\)
and \(\hat{s}=1.277788\). The scatter plot shows the Mazur fit in a red
dashed line and the Rachlin fit in a solid blue line.

\hypertarget{section-2}{%
\subsection{(3)}\label{section-2}}

\[
\begin{aligned}
\frac{A}{2} &= \frac{A}{1+kD^s},\text{ divide both sides by A, then cross multiply} \\
2 &= 1 + kD^s,\text{ subtract 1 from both sides}\\
1 &= kD^s,\text{ divide both sides by k, then exponentiate both sides to the 1/s power}\\
\Big(\frac{1}{k}\Big)^{1/s} &=D
\end{aligned}
\] Thus, the delay \(D\) at which the regression line is half the larger
later value is \(\Big(\frac{1}{k}\Big)^{1/s}\). This is the general
process for obtaining analytical solutions for ED50, see (Franck et al.
2015) for more details about ED50 for different discounting models.

\hypertarget{section-3}{%
\subsection{(4)}\label{section-3}}

\begin{Shaded}
\begin{Highlighting}[]
\NormalTok{ED50.Rachlin}\OtherTok{=}\NormalTok{(}\DecValTok{1}\SpecialCharTok{/}\NormalTok{k.hat)}\SpecialCharTok{\^{}}\NormalTok{(}\DecValTok{1}\SpecialCharTok{/}\NormalTok{s.hat)}
\NormalTok{ED50.Rachlin}
\end{Highlighting}
\end{Shaded}

\begin{verbatim}
## [1] 1415.088
\end{verbatim}

For the Mazur model, the estimated ED50 value is
\(\frac{1}{\hat{k}}=1418\) days (from earlier in chapter). For the
Rachlin model, the estimated ED50 is
\(\Big(\frac{1}{\hat{k}}\Big)^{1/\hat{s}}=1415\) days. It is
unsurprising that these values are close, because as the above plot
shows, the lines happen to be very close to each other when indifference
point =0.5. We do not expect ED90 values to agree very much, because at
an indifference point value of 0.1, these curves are further apart.

\hypertarget{section-4}{%
\subsection{(5)}\label{section-4}}

\begin{Shaded}
\begin{Highlighting}[]
\FunctionTok{table}\NormalTok{(dat}\SpecialCharTok{$}\NormalTok{JBviol)}
\end{Highlighting}
\end{Shaded}

\begin{verbatim}
## 
##  0  1 
## 82 24
\end{verbatim}

\begin{Shaded}
\begin{Highlighting}[]
\FunctionTok{table}\NormalTok{(dat}\SpecialCharTok{$}\NormalTok{ddattend)}
\end{Highlighting}
\end{Shaded}

\begin{verbatim}
## 
##       $0.00  now $100.00 in 1 day 
##                6              100
\end{verbatim}

\begin{Shaded}
\begin{Highlighting}[]
\FunctionTok{table}\NormalTok{(dat}\SpecialCharTok{$}\NormalTok{JBviol,dat}\SpecialCharTok{$}\NormalTok{ddattend)}
\end{Highlighting}
\end{Shaded}

\begin{verbatim}
##    
##     $0.00  now $100.00 in 1 day
##   0          1               81
##   1          5               19
\end{verbatim}

Twenty four participants violate the Johnson \& Bickel criteria. Six
individuals failed the attention check by answering they would rather
have zero dollars now instead of one hundred dollars in a day. Five of
the six individuals who failed the attention check also violated Johnson
\& Bickel criteria.

\hypertarget{section-5}{%
\subsection{(6)}\label{section-5}}

\begin{Shaded}
\begin{Highlighting}[]
\NormalTok{dat.noJB}\OtherTok{\textless{}{-}}\NormalTok{dat[dat}\SpecialCharTok{$}\NormalTok{JBviol}\SpecialCharTok{==}\DecValTok{0}\NormalTok{,]}
\end{Highlighting}
\end{Shaded}

The above code creates a data set called dat.noJB which starts as the
original dat set, then (using square bracket notation), only rows with
JBbiol=0 are retained (since this condition is before the comma in the
square brackets). Since there is no condition after the comma, this
indicates that all columns from dat remain in dat.noJB. Complete the
analysis by using the new data object to perform similar steps to what
we saw in the chapter.

\hypertarget{section-6}{%
\subsection{(7)}\label{section-6}}

\begin{Shaded}
\begin{Highlighting}[]
\FunctionTok{library}\NormalTok{(pracma)}
\NormalTok{AUC.vec}\OtherTok{\textless{}{-}}\FunctionTok{c}\NormalTok{() }\CommentTok{\#initialize AUC.vec as an empty vector}
\ControlFlowTok{for}\NormalTok{(i }\ControlFlowTok{in} \DecValTok{1}\SpecialCharTok{:}\DecValTok{106}\NormalTok{)\{ }\CommentTok{\#begin at i=1, do everything in curly brackets, increment i, repeat}
\NormalTok{  y.frame}\OtherTok{\textless{}{-}}\NormalTok{dat[i,}\DecValTok{5}\SpecialCharTok{:}\DecValTok{11}\NormalTok{,drop}\OtherTok{=}\ConstantTok{FALSE}\NormalTok{] }\CommentTok{\#pull indifference points for ith participant}
\NormalTok{  y}\OtherTok{\textless{}{-}}\FunctionTok{as.vector}\NormalTok{(}\FunctionTok{as.matrix}\NormalTok{(}\FunctionTok{t}\NormalTok{(y.frame)))  }\CommentTok{\#turn data frame y.frame into vector y}
\NormalTok{  AUC.vec[i]}\OtherTok{\textless{}{-}}\FunctionTok{trapz}\NormalTok{(D,y)}
\NormalTok{\}}
\NormalTok{ln.AUC.vec}\OtherTok{\textless{}{-}}\FunctionTok{log}\NormalTok{(AUC.vec)}
\FunctionTok{hist}\NormalTok{(AUC.vec)}
\end{Highlighting}
\end{Shaded}

\includegraphics{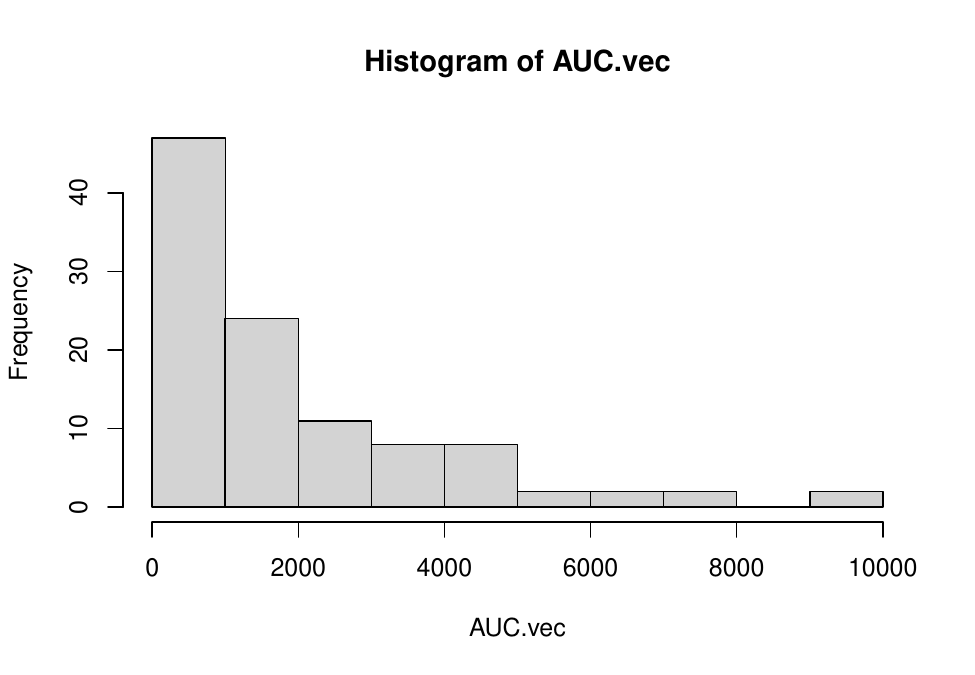}

\begin{Shaded}
\begin{Highlighting}[]
\FunctionTok{hist}\NormalTok{(ln.AUC.vec)}
\end{Highlighting}
\end{Shaded}

\includegraphics{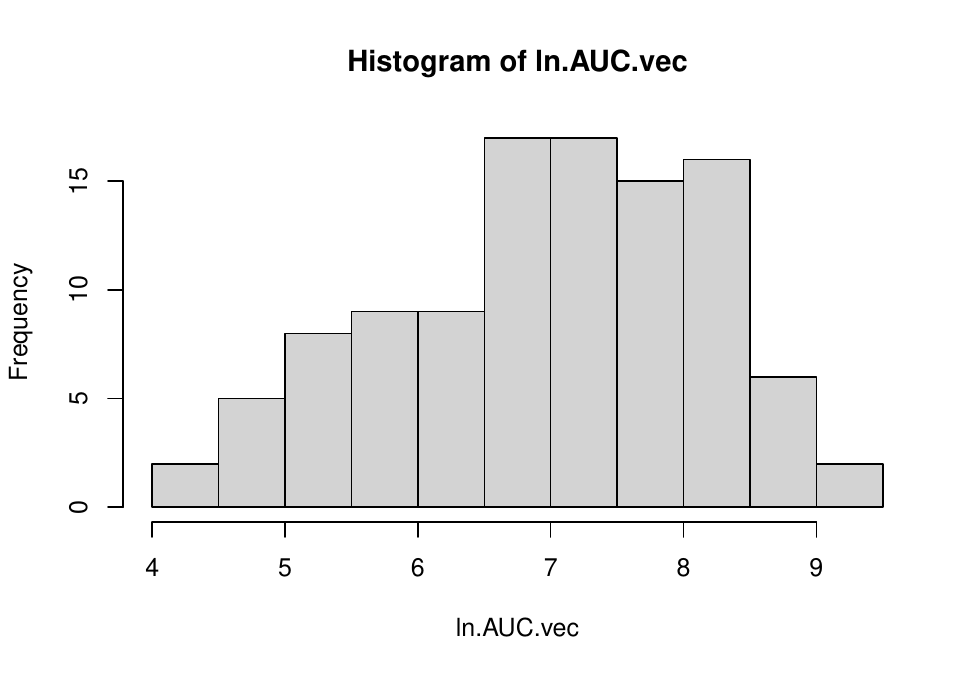}
Neither really looks like a normal distribution.

\hypertarget{section-7}{%
\subsection{(8)}\label{section-7}}

\begin{Shaded}
\begin{Highlighting}[]
\FunctionTok{library}\NormalTok{(ggplot2)}
\FunctionTok{library}\NormalTok{(GGally)}
\NormalTok{ED50.Mazur}\OtherTok{\textless{}{-}}\DecValTok{1}\SpecialCharTok{/}\NormalTok{K.vec}
\NormalTok{plot.frame}\OtherTok{\textless{}{-}}\FunctionTok{data.frame}\NormalTok{(}\AttributeTok{lnk=}\NormalTok{lnk,}\AttributeTok{ED50=}\NormalTok{ED50.Mazur,}\AttributeTok{AUC=}\NormalTok{AUC.vec,}\AttributeTok{ln.AUC=}\NormalTok{ln.AUC.vec)}
\NormalTok{plot.ob }\OtherTok{\textless{}{-}} \FunctionTok{ggpairs}\NormalTok{(}
\NormalTok{  plot.frame,}
  \AttributeTok{upper =} \FunctionTok{list}\NormalTok{(}\AttributeTok{continuous =} \StringTok{"density"}\NormalTok{, }\AttributeTok{combo =} \StringTok{"box\_no\_facet"}\NormalTok{),}
  \AttributeTok{lower =} \FunctionTok{list}\NormalTok{(}\AttributeTok{continuous =} \StringTok{"points"}\NormalTok{, }\AttributeTok{combo =} \StringTok{"dot\_no\_facet"}\NormalTok{)}
\NormalTok{)}
\NormalTok{plot.ob}
\end{Highlighting}
\end{Shaded}

\includegraphics{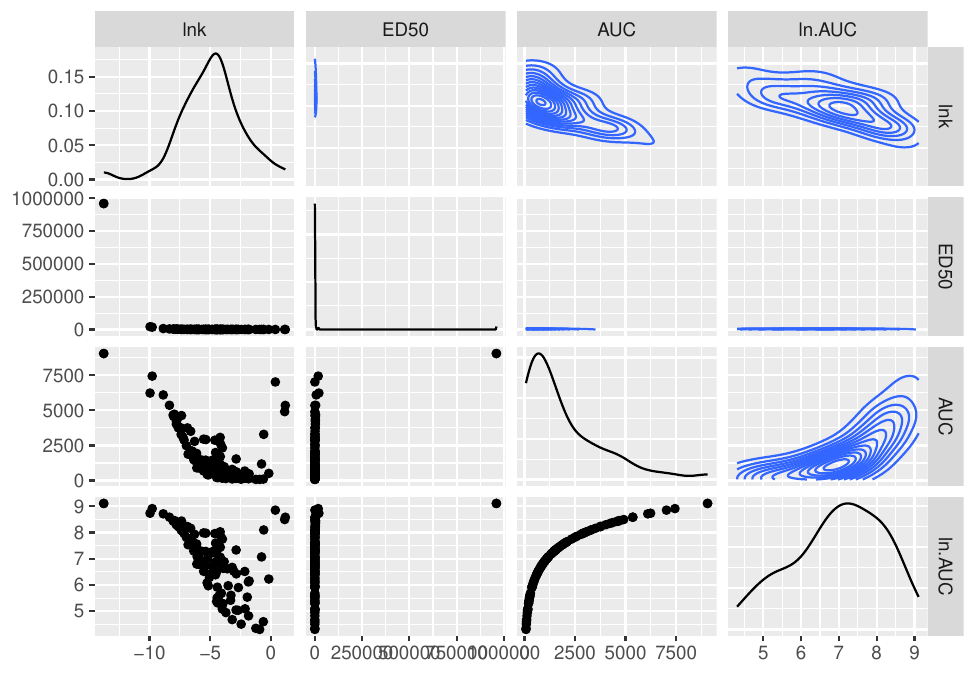}

The above plot visualizes the distributions of and associations among
four discouting metrics, but does not definitively recommend any as
intrinsically better than the others. We note an outlier in lnk, and we
might consider log-transforming ED50 in a revision of this plot.

\hypertarget{section-8}{%
\subsection{(9)}\label{section-8}}

\begin{Shaded}
\begin{Highlighting}[]
\NormalTok{mod}\OtherTok{\textless{}{-}}\FunctionTok{lm}\NormalTok{(lnk}\SpecialCharTok{\textasciitilde{}}\NormalTok{dat}\SpecialCharTok{$}\NormalTok{age)}
\NormalTok{preds}\OtherTok{\textless{}{-}}\FunctionTok{predict}\NormalTok{(mod)}
\NormalTok{resids}\OtherTok{\textless{}{-}}\FunctionTok{resid}\NormalTok{(mod)}
\FunctionTok{hist}\NormalTok{(resids) }\CommentTok{\#Eh good enough}
\end{Highlighting}
\end{Shaded}

\includegraphics{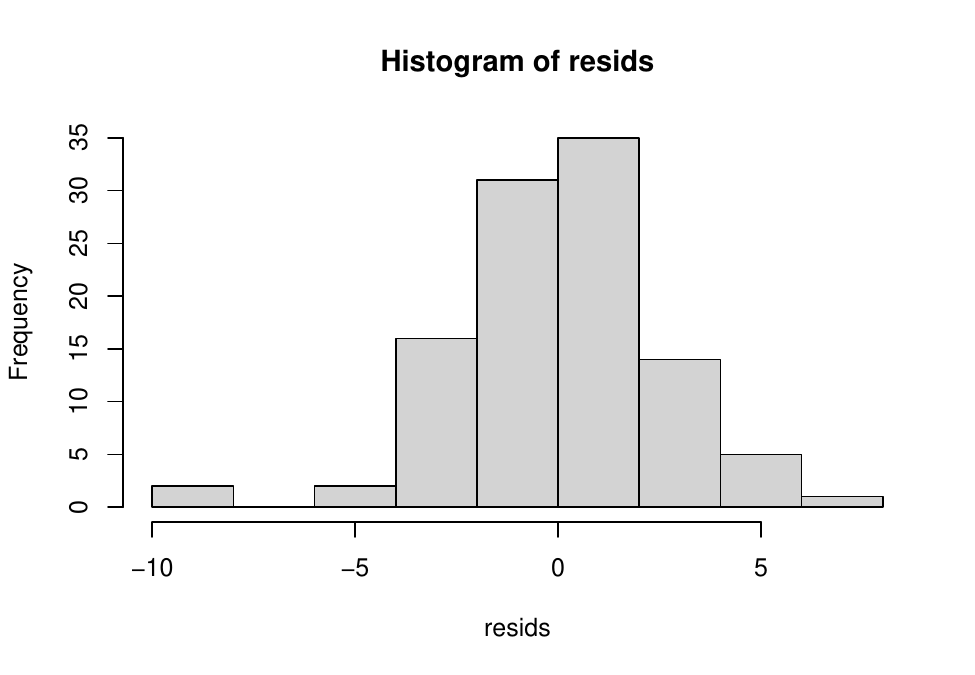}

\begin{Shaded}
\begin{Highlighting}[]
\FunctionTok{plot}\NormalTok{(preds, resids) }\CommentTok{\#*Maybe* some non constant variance}
\end{Highlighting}
\end{Shaded}

\includegraphics{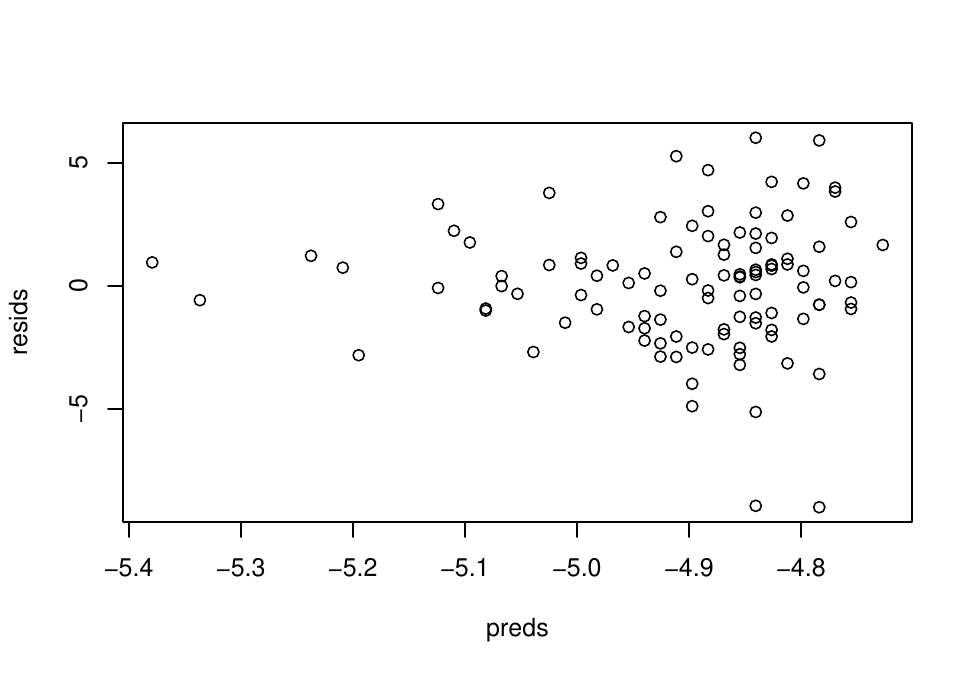}

\begin{Shaded}
\begin{Highlighting}[]
\CommentTok{\#probably ok}
\end{Highlighting}
\end{Shaded}

Both assumptions seem to be reasonably well met, though neither is
perfect. I'd personally be comfortable analyzing the data with the
proposed model.

\hypertarget{references}{%
\subsection*{References}\label{references}}
\addcontentsline{toc}{subsection}{References}

\hypertarget{refs}{}
\begin{CSLReferences}{1}{0}
\leavevmode\vadjust pre{\hypertarget{ref-Allison2001missing}{}}%
Allison, P. D. 2001. \emph{Missing Data}. Quantitative Applications in
the Social Sciences. SAGE Publications.
\url{https://books.google.com/books?id=LJB2AwAAQBAJ}.

\leavevmode\vadjust pre{\hypertarget{ref-Benjamin2018}{}}%
Benjamin, Daniel J., James O. Berger, Magnus Johannesson, Brian A.
Nosek, E.-J. Wagenmakers, Richard Berk, Kenneth A. Bollen, et al. 2018.
{``Redefine Statistical Significance.''} \emph{Nature Human Behaviour} 2
(1): 6--10. \url{https://doi.org/10.1038/s41562-017-0189-z}.

\leavevmode\vadjust pre{\hypertarget{ref-Borges2016}{}}%
Borges, Allison M., Jinyi Kuang, Hannah Milhorn, and Richard Yi. 2016.
{``An Alternative Approach to Calculating {Area}-{Under}-the-{Curve}
({AUC}) in Delay Discounting Research.''} \emph{Journal of the
Experimental Analysis of Behavior} 106 (2): 145--55.
\url{https://doi.org/10.1002/jeab.219}.

\leavevmode\vadjust pre{\hypertarget{ref-Chavez2017hierarchical}{}}%
Chávez, Melisa E, Elena Villalobos, José L Baroja, and Arturo Bouzas.
2017. {``Hierarchical Bayesian Modeling of Intertemporal Choice.''}
\emph{Judgment and Decision Making} 12 (1): 19--28.

\leavevmode\vadjust pre{\hypertarget{ref-Open2015estimating}{}}%
Collaboration, Open Science. 2015. {``Estimating the Reproducibility of
Psychological Science.''} \emph{Science} 349 (6251): aac4716.

\leavevmode\vadjust pre{\hypertarget{ref-Franck2015accurate}{}}%
Franck, Christopher T., Mikhail N. Koffarnus, Leanna L. House, and
Warren K. Bickel. 2015. {``Accurate Characterization of Delay
Discounting: A Multiple Model Approach Using Approximate Bayesian Model
Selection and a Unified Discounting Measure.''} \emph{Journal of the
Experimental Analysis of Behavior} 103 (1): 218--33.
https://doi.org/\url{https://doi.org/10.1002/jeab.128}.

\leavevmode\vadjust pre{\hypertarget{ref-Franck2019overview}{}}%
Franck, Christopher T., Mikhail N. Koffarnus, Todd L. McKerchar, and
Warren K. Bickel. 2019. {``An Overview of {Bayesian} Reasoning in the
Analysis of Delay-Discounting Data.''} \emph{Journal of the Experimental
Analysis of Behavior} 111 (2): 239--51.
\url{https://doi.org/10.1002/jeab.504}.

\leavevmode\vadjust pre{\hypertarget{ref-Franck2022how}{}}%
Franck, Christopher T., Michael L. Madigan, and Nicole A. Lazar. 2022.
{``How to Write about Alternatives to Classical Hypothesis Testing
Outside of the Statistical Literature: Approximate Bayesian Model
Selection Applied to a Biomechanics Study.''} \emph{Stat} 11 (1): e508.
https://doi.org/\url{https://doi.org/10.1002/sta4.508}.

\leavevmode\vadjust pre{\hypertarget{ref-Franck2023tribute}{}}%
Franck, Christopher T., Haily K. Traxler, Brent A. Kaplan, Mikhail N.
Koffarnus, and Mark J. Rzeszutek. 2023. {``A Tribute to {Howard}
{Rachlin} and His Two-Parameter Discounting Model: {Reliable} and
Flexible Model Fitting.''} \emph{Journal of the Experimental Analysis of
Behavior} 119 (1): 156--68. \url{https://doi.org/10.1002/jeab.820}.

\leavevmode\vadjust pre{\hypertarget{ref-Gilroy2018discounting}{}}%
Gilroy, Shawn P., and Donald A. Hantula. 2018. {``Discounting Model
Selection with Area-Based Measures: A Case for Numerical Integration.''}
\emph{Journal of the Experimental Analysis of Behavior} 109 (2):
433--49. https://doi.org/\url{https://doi.org/10.1002/jeab.318}.

\leavevmode\vadjust pre{\hypertarget{ref-Ioannidis2005why}{}}%
Ioannidis, John P. A. 2005. {``Why Most Published Research Findings Are
False.''} \emph{PLoS Medicine} 2 (8): e124.
\url{https://doi.org/10.1371/journal.pmed.0020124}.

\leavevmode\vadjust pre{\hypertarget{ref-Cohen2007power}{}}%
Jacob, Cohen. 2007. {``A Power Primer.''} \emph{Tutorials in
Quantitative Methods for Psychology} 112 (September).
\url{https://doi.org/10.1037/0033-2909.112.1.155}.

\leavevmode\vadjust pre{\hypertarget{ref-Johnson2008algorithm}{}}%
Johnson, Matthew W., and Warren K. Bickel. 2008. {``An Algorithm for
Identifying Nonsystematic Delay-Discounting Data.''} \emph{Experimental
and Clinical Psychopharmacology} 16 (3): 264--74.
\url{https://doi.org/10.1037/1064-1297.16.3.264}.

\leavevmode\vadjust pre{\hypertarget{ref-Kaplan2021applying}{}}%
Kaplan, Brent A., Christopher T. Franck, Kevin McKee, Shawn P. Gilroy,
and Mikhail N. Koffarnus. 2021. {``Applying {Mixed}-{Effects} {Modeling}
to {Behavioral} {Economic} {Demand}: {An} {Introduction}.''}
\emph{Perspectives on Behavior Science} 44 (2): 333--58.
\url{https://doi.org/10.1007/s40614-021-00299-7}.

\leavevmode\vadjust pre{\hypertarget{ref-Killeen2023variations}{}}%
Killeen, Peter R. 2023. {``Variations on a Theme by Rachlin: Probability
Discounting.''} \emph{Journal of the Experimental Analysis of Behavior}
119 (1): 140--55.
https://doi.org/\url{https://doi.org/10.1002/jeab.817}.

\leavevmode\vadjust pre{\hypertarget{ref-Koffarnus2014a5trial}{}}%
Koffarnus, Mikhail N., and Warren K. Bickel. 2014. {``A 5-Trial
Adjusting Delay Discounting Task: Accurate Discount Rates in Less Than
One Minute.''} \emph{Experimental and Clinical Psychopharmacology} 22
(3): 222--28. \url{https://doi.org/10.1037/a0035973}.

\leavevmode\vadjust pre{\hypertarget{ref-Leek2015pvalues}{}}%
Leek, Jeffrey T., and Roger D. Peng. 2015. {``Statistics: {P} Values Are
Just the Tip of the Iceberg.''} \emph{Nature} 520 (7549): 612--12.
\url{https://doi.org/10.1038/520612a}.

\leavevmode\vadjust pre{\hypertarget{ref-Little2014statistical}{}}%
Little, R. J. A., and D. B. Rubin. 2014. \emph{Statistical Analysis with
Missing Data}. Wiley Series in Probability and Statistics. Wiley.
\url{https://books.google.com/books?id=AyVeBAAAQBAJ}.

\leavevmode\vadjust pre{\hypertarget{ref-Lohr2021sampling}{}}%
Lohr, S. L. 2021. \emph{Sampling: Design and Analysis}. Chapman \&
Hall/CRC Texts in Statistical Science. CRC Press.
\url{https://books.google.com/books?id=DahGEAAAQBAJ}.

\leavevmode\vadjust pre{\hypertarget{ref-Mazur1987adjusting}{}}%
Mazur, James E. 1987. {``An Adjusting Procedure for Studying Delayed
Reinforcement.''} In \emph{The Effect of Delay and of Intervening Events
on Reinforcement Value.}, 55--73. Quantitative Analyses of Behavior,
{Vol}. 5. Hillsdale, NJ, US: Lawrence Erlbaum Associates, Inc.

\leavevmode\vadjust pre{\hypertarget{ref-McKerchar2009comparison}{}}%
McKerchar, Todd, Leonard Green, Joel Myerson, T. Pickford, Jade Hill,
and Steven Stout. 2009. {``A Comparison of Four Models of Delay
Discounting in Humans.''} \emph{Behavioural Processes} 81 (June):
256--59. \url{https://doi.org/10.1016/j.beproc.2008.12.017}.

\leavevmode\vadjust pre{\hypertarget{ref-Montgomery2008design}{}}%
Montgomery, D. C. 2008. \emph{Design and Analysis of Experiments}.
Design and Analysis of Experiments. John Wiley \& Sons.
\url{https://books.google.com/books?id=kMMJAm5bD34C}.

\leavevmode\vadjust pre{\hypertarget{ref-Montgomery2015introduction}{}}%
Montgomery, D. C., E. A. Peck, and G. G. Vining. 2015.
\emph{Introduction to Linear Regression Analysis}. Wiley Series in
Probability and Statistics. Wiley.
\url{https://books.google.ca/books?id=27kOCgAAQBAJ}.

\leavevmode\vadjust pre{\hypertarget{ref-Myerson2001area}{}}%
Myerson, J., L. Green, and M. Warusawitharana. 2001. {``Area Under the
Curve as a Measure of Discounting.''} \emph{Journal of the Experimental
Analysis of Behavior} 76 (2): 235--43.
\url{https://doi.org/10.1901/jeab.2001.76-235}.

\leavevmode\vadjust pre{\hypertarget{ref-Odum2011delay}{}}%
Odum, Amy L. 2011. {``DELAY DISCOUNTING: I'm a k, YOU'RE a k.''}
\emph{Journal of the Experimental Analysis of Behavior} 96 (3): 427--39.
https://doi.org/\url{https://doi.org/10.1901/jeab.2011.96-423}.

\leavevmode\vadjust pre{\hypertarget{ref-Odum2020delay}{}}%
Odum, Amy L., Ryan J. Becker, Jeremy M. Haynes, Ann Galizio, Charles C.
J. Frye, Haylee Downey, Jonathan E. Friedel, and D. M. Perez. 2020.
{``Delay Discounting of Different Outcomes: Review and Theory.''}
\emph{Journal of the Experimental Analysis of Behavior} 113 (3):
657--79. https://doi.org/\url{https://doi.org/10.1002/jeab.589}.

\leavevmode\vadjust pre{\hypertarget{ref-RCoreTeam2022R}{}}%
R Core Team. 2022. \emph{R: A Language and Environment for Statistical
Computing}. Vienna, Austria: R Foundation for Statistical Computing.
\url{https://www.R-project.org/}.

\leavevmode\vadjust pre{\hypertarget{ref-Rachlin2006notes}{}}%
Rachlin, Howard. 2006. {``Notes on Discounting.''} \emph{Journal of the
Experimental Analysis of Behavior} 85 (June): 425--35.
\url{https://doi.org/10.1901/jeab.2006.85-05}.

\leavevmode\vadjust pre{\hypertarget{ref-Rachlin1986cognition}{}}%
Rachlin, Howard, Alexandra Logue, John Gibbon, and Marvin Frankel. 1986.
{``Cognition and Behavior in Studies of Choice.''} \emph{Psychological
Review} 93 (January): 33--45.
\url{https://doi.org/10.1037/0033-295X.93.1.33}.

\leavevmode\vadjust pre{\hypertarget{ref-Rachlin1991subjective}{}}%
Rachlin, Howard, Andres Raineri, and David Cross. 1991. {``SUBJECTIVE
PROBABILITY AND DELAY.''} \emph{Journal of the Experimental Analysis of
Behavior} 55 (2): 233--44.
https://doi.org/\url{https://doi.org/10.1901/jeab.1991.55-233}.

\leavevmode\vadjust pre{\hypertarget{ref-Traxler2022interest}{}}%
Traxler, Haily K., Brent A. Kaplan, Mark J. Rzeszutek, Christopher T.
Franck, and Mikhail N. Koffarnus. 2022. {``Interest in and Perceived
Effectiveness of Contingency Management Among Alcohol Drinkers Using
Behavioral Economic Purchase Tasks.''} \emph{Experimental and Clinical
Psychopharmacology}, No Pagination Specified--.
\url{https://doi.org/10.1037/pha0000580}.

\leavevmode\vadjust pre{\hypertarget{ref-Utts2014mind}{}}%
Utts, J. M., and R. F. Heckard. 2014. \emph{Mind on Statistics}. Cengage
Learning. \url{https://books.google.com/books?id=PuLKAgAAQBAJ}.

\leavevmode\vadjust pre{\hypertarget{ref-Wackerly2014mathematical}{}}%
Wackerly, D., W. Mendenhall, and R. L. Scheaffer. 2014.
\emph{Mathematical Statistics with Applications}. Cengage Learning.
\url{https://books.google.ca/books?id=lTgGAAAAQBAJ}.

\leavevmode\vadjust pre{\hypertarget{ref-Wasserstein2016asa}{}}%
Wasserstein, Ronald L., and Nicole A. Lazar. 2016. {``The ASA Statement
on p-Values: Context, Process, and Purpose.''} \emph{The American
Statistician} 70 (2): 129--33.
\url{https://doi.org/10.1080/00031305.2016.1154108}.

\leavevmode\vadjust pre{\hypertarget{ref-Yoon2008turning}{}}%
Yoon, Jin H., and Stephen T. Higgins. 2008. {``Turning k on Its Head:
Comments on Use of an {ED50} in Delay Discounting Research.''}
\emph{Drug and Alcohol Dependence} 95 (1-2): 169--72.
\url{https://doi.org/10.1016/j.drugalcdep.2007.12.011}.

\leavevmode\vadjust pre{\hypertarget{ref-Young2017discounting}{}}%
Young, Michael E. 2017. {``Discounting: {A} Practical Guide to
Multilevel Analysis of Indifference Data.''} \emph{Journal of the
Experimental Analysis of Behavior} 108 (1): 97--112.
\url{https://doi.org/10.1002/jeab.265}.

\leavevmode\vadjust pre{\hypertarget{ref-Young2011deming}{}}%
Young, S Stanley, and Alan Karr. 2011. {``Deming, Data and Observational
Studies: A Process Out of Control and Needing Fixing.''}
\emph{Significance} 8 (3): 116--20.

\end{CSLReferences}

\end{document}